\documentclass[twocolumn,showpacs,nofootinbib]{revtex4}
\usepackage[latin1]{inputenc}
\usepackage{amssymb,eucal,amsthm,epsf,graphics,float}
\usepackage[english]{babel}
\renewcommand{\thesubsection}{\alph{subsection}}

\def\e{{\mathop{\,\rm e}}}
\def\ii{\dot{\i\hspace{-1.7pt}\i}}
\def\Z{\mathbb{Z}}
\def\circled#1{\bigcirc\hspace{-1.7ex}#1}
\def\notC{{\mathop{\not{\!C}}}}

\begin{document}
\title{Renormalization Group calculations with $k_\Vert$ dependent
couplings in a ladder}
\author{
G. Abramovici,
J. C. Nickel
and M. H\'eritier
}
\affiliation{Laboratoire de Physique des Solides, associé au C.N.R.S.,
Université de Paris Sud, Centre d'Orsay 91405 Orsay France}
\begin{abstract}
We calculate the phase diagram of a ladder system, with a Hubbard interaction
and an interchain coupling $t_\perp$. We use a Renormalization
Group method, in a one loop expansion, introducing an original method to include
$k_\Vert$ dependence of couplings. We also classify the order parameters
corresponding to ladder instabilities. We obtain different results, depending
on whether we include $k_\Vert$ dependence or not. When we do so, we observe
a region with large antiferromagnetic fluctuations, in the vicinity of small
$t_\perp$, followed by a superconducting region with a simultaneous divergence
of the Spin Density Waves channel. We also investigate
the effect of a non local backward interchain scattering~: we observe, on one
hand, the suppression of singlet superconductivity and of Spin Density Waves,
and, on the other hand, the increase of Charge Density Waves and, for some
values of $t_\perp$, of triplet superconductivity. Our results eventually show
that $k_\Vert$ is an influential variable in the Renormalization Group
flow, for this kind of systems.
\end{abstract}
\maketitle

\section{Introduction}

The physics of ladder systems remains a source of considerable interest. In the
last decades, many conductors were found, with anisotropic two-leg
electronic structure, such as $Sr Cu_2 O_3$\cite{Azuma} or
$Sr_{14-x} Ca_x Cu_{24} O_{41}$\cite{Carron} compounds. The structure of $La_2
Cu_2 O_5$\cite{Hiroi} can also be analyzed as weakly coupled ladders, and is
therefore very similar. The phase diagram of these compounds is very rich;
it is now well established that these systems behave like
Luttinger liquids at high temperature, while they can behave like Fermi
liquids when $T$ decreases; they exhibit superconducting (SC) phases of type
II, which can also be mixed with antiferromagnetic fluctuations. In some case,
the SC phase is found to be spin-gapped, while spinless phases are also
reported\cite{Dagotto}.

From a theoretical point of view, the ladder (or two-coupled chain) model is the
simplest quasi-one-dimensional one.  Although all its properties are not
entirely elucidated, it has been used by many authors as a toy-model, to build
and explore new approaches (two-leaf dispersion models  calculated by a
Kadanoff-Wilson renormalization method\cite{Penc} or by a bosonization
method\cite{Giamarchi}; transition between commensurate and incommensurate
filling\cite{Patrick}, in particular close to the half-filing
case\cite{Kishine,Karine}; dimensional transition\cite{Sonia}, etc.).  These
systems have been intensively used to investigate non conventional SC process
(with singlet\cite{Dagotto1,Fabrizio,Khveshchenko} or triplet\cite{Barnes}
pairing).

This paper is devoted to the study of a ladder model, with a Hubbard interaction
($U$ is the Hubbard constant) and an interband coupling ($t_\perp$ is the
interaction factor), at zero temperature.  We investigate the phase diagram in
the range of parameter $0\le U\le8\pi t_\Vert\eta$ and suppose that the
electronic filling $\eta$ is incommensurate.  Although there is yet no evidence
for the existence of compounds lying in this range of parameter, we have
hopes that some of these will indeed be found and confirm the theoretical
predictions that we present here.

The phase diagram of this system has been partially studied by several authors.
In particular, M. Fabrizio have used a Renormalization Group (RG) method, in a
two-loop expansion from the Fermi liquid solution\cite{Fabrizio}.  He includes
interband backward scattering $g_b$, and, within the range of parameter that we
have investigated, finds a SC phase, named `phase I', which he clearly points
out not to be singlet $s$, although he did not elucidate the symmetry of the
condensate. In his calculations, the RG flow of susceptibilities shows several
divergences~: the Spin Density Wave (SDW) channel coexists with the
superconducting one. This author did alternatively  cut the flow before the
divergences and bozonize the effective Hamiltonian; then, SDW modes disappear
and are replaced by Charge Density Wave (CDW) instabilities.

These results are more or less compatible with that of other methods, using a
one-loop expansion\cite{Lin,Finkelstein}, or bozonizing the Hamiltonian of the
bare system\cite{Schulz,Orignac,Kuroki,Khveshchenko2} (see also
Ref.~\cite{Khveshchenko}), or using Quantum Monte Carlo method\cite{Scalapino}.
These authors generally find a singlet superconducting condensate of symmetry
$d$, which coexists with SDW or CDW instabilities.  This complicated phase has
also been related to the RVB phase\cite{Noack}.  One of the central question is
whether the SC modes are spin-gapped or not, and receives various and even
contradictory answers.  Using a two-loop expansion, Kishine\cite{Kishine2}
observes a spin gap, which is suppressed when $t_\perp\to0$.  This is also the
result found with a Density Matrix RG method by Park\cite{Park}.

We give a new insight into these questions, using a RG method, in a perturbative
expansion in $U$. This kind of method has been used in many recent and very
complete works\cite{Metzner,Honerkamp,Halboth}. Here, we calculate RG equations
with one loop diagrams, including $g_b$ couplings, as in Ref.~\cite{Fabrizio},
as well as all parallel momentum dependence.

Let us emphasize that, although we begin from the Fermi liquid solution, we find
a phase with only SDW fluctuations, for small enough interband interaction
$t_\perp\le t_{\perp c}(U)$, contrary to all previous results obtained by RG

methods, which always indicate SC instabilities as soon as $t_\perp\ne0$ (see
for instance Lin {\it et al.}\cite{Lin}).
However, this phase is different from the one-dimension limit.  This is a very
remarkable result, since $k_\Vert$ is known to be an irrelevant variable
of the RG flow\cite{Shankar}. However, we will prove in this paper that it is
indeed influential in the very case of a ladder.

When $t_\perp$ is increased, we observe a  transition at $t_{\perp c}$ to a
superconducting phase, where singlet SC instabilities of symmetry $d$ coexist
with SDW ones. This phase is found by many authors (see Refs.~\cite{Fabrizio,
Lin,Finkelstein} and other articles already quoted).

Recently, Bourbonnais {\it et al.}\cite{Bourbonnais2,Nickel2} have examined the
effect of interchain Coulomb interactions for an infinite number of coupled
chains, using RG method. Interchain backward scattering was found to enhance
CDW fluctuations and favor triplet instead of singlet SC. We here investigated
the effect of a Coulomb interchain backward interaction $C_{\rm back}$ for the
ladder problem. We find that this interaction favors triplet SC instabilities
instead of singlet ones and CDW instabilities instead of SDW ones in a ladder
system too. Indeed, both the singlet SC and SDW instabilities (if any) are
suppressed when $C_{\rm back}$ is increased, and we observe triplet SC as well
as CDW instabilities. The triplet SC existence region is however very narrow and
lies inside the region $t_\perp\ge t_{\perp c}(U)$. For large values of
$C_{\rm back}$ ($C_{\rm back}\sim U$), CDW is always dominant; this is
consistent with what Lin {\it et al.} find\cite{Lin}.

On the contrary, when a Coulombian interchain forward interaction $C_{\rm for}$
is added, all SC instabilities are depressed, and we only observe SDW and
CDW fluctuations.

We also present a detailed classification of the pair operators in a ladder,
which are connected to the order parameters. It proves
a very powerful tool in these sophisticated RG methods.

So, we will first give a short description of the model (in section {\bf 2}), then
present the classification of the pair operators (in section {\bf 3}, the
symmetries are explained in appendix~\ref{symmetries}), then we explain the
RG formulation and techniques that are used here (in section {\bf 4}). Results
concerning only initially local interactions are presented in section {\bf 5},
while those concerning the influence of additional interchain interactions are
given in section {\bf 6}. In section {\bf 7} we conclude.

\section{Model}

The Hubbard model of a ladder has been studied in various articles. We give here
a brief presentation of this model (see Refs.~\cite{Fabrizio} or \cite{Dusuel}
with similar notations).

\subsection{Kinetic Hamiltonian}

\subsubsection{The model in a 1-$d$ representation}

\begin{figure}[H]
\begin{center}
\begin{picture}(120,50)
\put(0,15){\vector(1,0){120}}
\put(60,10){\vector(0,1){25}}
\put(115,5){\makebox(0,0){$k$}}
\put(52,30){\makebox(0,0){$\epsilon_k$}}
\thicklines
\put(5,30){\line(1,-1){30}}
\put(17,30){\line(1,-1){30}}
\put(103,30){\line(-1,-1){30}}
\put(115,30){\line(-1,-1){30}}
\end{picture}
\end{center}
\caption{The 2-band dispersion in $\Vert$ direction}
\label{dispersion}
\end{figure}
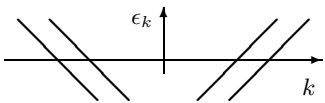

The dispersion curve separates into two bands (0 and $\pi$), so the Fermi
surface splits into four points ($-k_{f0}$, $-k_{f\pi}$, $k_{f\pi}$, $k_{f0}$ in
the $\Vert$ direction, + corresponds to right moving particles and $-$ to
left moving ones). The bands are linearized around the Fermi
vectors\cite{Solyom} with a single Fermi velocity $v_f$ (cf.
Fig.~\ref{dispersion}). We write $R$ the right moving particles and $L$ the
left moving ones. Then, the kinetic Hamiltonian writes
\begin{widetext}
\begin{eqnarray}
H_{\rm cin}&=&\sum_\sigma v_f\Big(\sum_K(K-k_{f0})R^\dag_{0\sigma}(K)R_{0\sigma}(K)+
\sum_K(K-k_{f\pi})R^\dag_{\pi\sigma}(K)R_{\pi\sigma}(K)\nonumber\\
&&+\sum_K(K+k_{f0})L^\dag_{0\sigma}(K)L_{0\sigma}(K)+
\sum_K(K+k_{f\pi})L^\dag_{\pi\sigma}(K)L_{\pi\sigma}(K)\Big).
\label{Hamiltoncin}
\end{eqnarray}
\end{widetext}
We define the Fermi surface gap $\Delta k_f=k_{f0}-k_{f\pi}$.  One then gets
$\Delta k_f=2t_\perp/v_f$. The discretization step in $\Vert$ direction is
$a$, and the reciprocal vector in this direction is defined modulo $2\pi/a$.
The distance between the chains in $\perp$ direction is $b$.

In order to give a clear representation of all instability processes that will
be discussed afterwards, it is worth showing how this model can be embedded in a
2-$d$ representation, which we present now.

\subsubsection{The model in a 2-$d$ representation}

\begin{figure}[t]
\epsfysize=10cm
\epsfbox{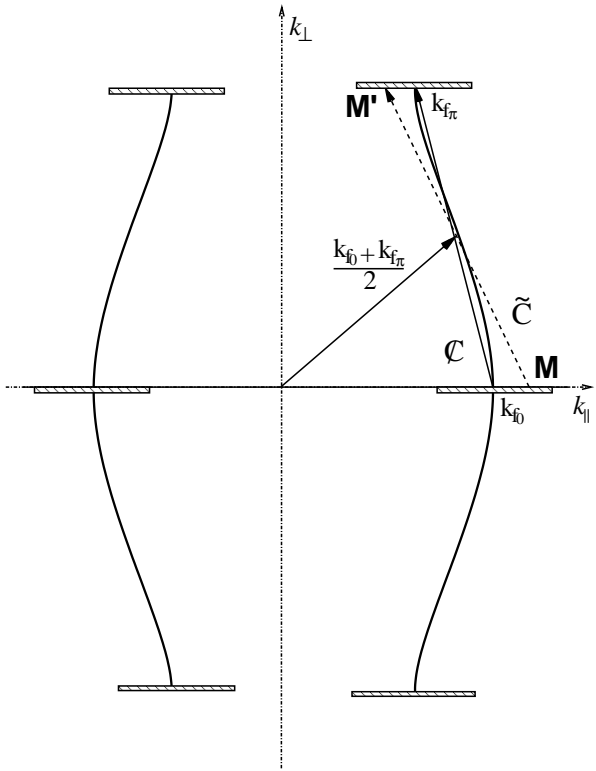}
\caption{Two-dimensional representation of the states.  The physical space, for
a ladder, is restricted to the hatched bands at $k_\perp=0$ and
$k_\perp=\pm{\pi\over b}$.
The curves correspond to the Fermi surface of a 2-$d$ system, with an infinite
number of chains.  The four Fermi points, in the ladder system, are the
intersections of these curves with the physical bands.  The symmetry $\tilde C$
that maps point $M$ onto $M'$ is the point symmetry around the affine point,
corresponding to $k_{f0}+k_{f\pi}\over2$.  The symmetry $\notC$ is the
translation by the vector $k_{f0}-k_{f\pi}$, which is also represented by a
plain arrow.}
\label{para2d}
\end{figure}

The general 2-$d$ dispersion law writes
$$
\epsilon({\bf k})=-2t_\Vert\cos(k_\Vert a)-2t_\perp\cos(k_\perp b)
 $$
as represented in Fig.~\ref{para2d}; if one writes approximately $k_f\approx
k_{f0}\approx k_{f\pi}$, one gets $v_f\approx2at_\Vert\sin(k_fa)$.
In real space, $y_\Vert$ corresponds
to the axis along the chains, and $y_\perp$ takes only two values, $y_\perp=
\pm1$, corresponding to which chain one refers to. The $\Vert$ axis is then
discretized, $y_\Vert=1\cdot\cdot N$, where $4N$ is the total number of
states. The Fourier transformation from real space functions to reciprocal
space ones is detailed in appendix~\ref{Fourier}.

From $y_\perp=\pm1$, one gets $k_\perp=0$ or $\pm{\pi\over b}$ ($k_\perp=
{\pi\over b}$ and $k_\perp=-{\pi\over b}$ are identified). Therefore, the
physical states are limited to the four horizontal bands, shown on
Fig.~\ref{para2d}, which are centered on each of the four Fermi points.
The two bands centered on $(\pm k_{f0},0)$ are the left and right bands 0
($L_0$ or $R_0$), the two bands centered on $(\pm k_{f\pi},{\pi\over b})$
are the left and right bands $\pi$ ($L_\pi$ or $R_\pi$).

\subsection{Interaction Hamiltonian}

The most general interaction Hamiltonian can be written, with
$L_{\bf k\sigma}=L_{0\sigma}(K)$ for ${\bf k}=(K,0)$,
$L_{\bf k\sigma}=L_{\pi\sigma}(K)$ for ${\bf k}=(K,\pi)$ and idem for $R$,
\begin{widetext}
\begin{eqnarray}\!\!\!\!\!\!\!\!
H_{\rm int}&=&{1\over N}\sum_{{\bf k}_1,{\bf k}_2,{\bf k}'_1,{\bf k}'_2
\atop{\bf k}_1+{\bf k}_2={\bf k}'_1+{\bf k}'_2}
\sum_{\sigma_1,\sigma_2}
{\cal G}_1({\bf k}_1,{\bf k}_2,{\bf k}'_2,{\bf k}'_1)
R_{{\bf k}'_1\sigma_1}^\dag L_{{\bf k}'_2\sigma_2}^\dag
R^{\phantom{\dag}}_{{\bf k}_1\sigma_1}
L^{\phantom{\dag}}_{{\bf k}_2\sigma_2}%\nonumber\\&&
+{\cal G}_2({\bf k}_1,{\bf k}_2,{\bf k}'_2,{\bf k}'_1)
R_{{\bf k}'_1\sigma_1}^\dag L_{{\bf k}'_2\sigma_2}^\dag
L^{\phantom{\dag}}_{{\bf k}_2\sigma_2}
R^{\phantom{\dag}}_{{\bf k}_1\sigma_1}
\label{Hamiltonint}
\end{eqnarray}
\end{widetext}
where ${\cal G}_\alpha$ is the two-particle coupling, and we have used the g-ology
representation. We do not include umklapp interactions ${\cal G}_3$($LLRR$ or
$RRLL$) nor ${\cal G}_4$ terms ($LLLL$ or $RRRR$). We will alternatively use
the singlet-triplet representation, where $\alpha$ takes values $\alpha=s,t$,
or the Charge-Spin representation, where $\alpha=C,S$. If not necessary, we will
omit the spin dependence $\alpha$.
Eqs.~(\ref{relgSC}) and (\ref{relgSDW}) give, in appendix~\ref{defcouplings},
the usual relations between these different representations.

We distinguish, following Fabrizio\cite{Fabrizio}, $g_0$,
 which\penalty-20000\ corresponds to
$R^\dag_0L^\dag_0L_0R_0$ process, $g_\pi$ ($\leftrightarrow$\penalty-20000\
$R^\dag_\pi L^\dag_\pi L_\pi R_\pi$), $g_{\!f0}$ ($\leftrightarrow$
$R^\dag_0L^\dag_\pi L_\pi R_0$), $g_{\!f\pi}$ ($\leftrightarrow$
$R^\dag_\pi L^\dag_0L_0R_\pi),\!\!\!\!\!\!\!\!$ $\quad g_{t0}$ ($\leftrightarrow$
$R^\dag_0L^\dag_0L_\pi R_\pi$), $g_{t\pi}$ ($\leftrightarrow$
$R^\dag_\pi L^\dag_\pi L_0R_0$), $g_{b0}$ ($\leftrightarrow$
$R^\dag_0L^\dag_\pi L_0R_\pi$) and $g_{b\pi}$ ($\leftrightarrow$
$R^\dag_\pi L^\dag_0L_\pi R_0$).  The definitions of these couplings, including
the $k_\Vert$ dependence are detailed in~\ref{defcouplings} and
Fig.~\ref{figcouplings}; from symmetry considerations, one only needs $g_0$,
$g_\pi$, $g_{\!f0}$, $g_{t0}$ and $g_{b0}$; in fact, we will see that even $g_\pi$
can be deduced from $g_0$ in the very case of a ladder, so that we only deal
with four couplings.

\paragraph*{Bare couplings}

Of course, in the initial Hubbard model, the two-particle couplings do not
depend on the momenta. We will define $\tilde U=U a/(\pi v_f)$ and
$\tilde g=g a/(\pi v_f)$, to get rid of the physical dimensions. Then, the
bare couplings values are simply $\tilde g_i=\tilde U$.

\paragraph*{Additional Coulombian interchain interactions}

The above Hamiltonian $H_{\rm int}$ corresponds to local interactions. We
have also studied the effect of additional Coulombian interchain interactions.

In order to implement a backward interchain interaction, we need to modulate
the bare parameters, which simply writes, in this case, $\tilde g_{01}=\tilde U
+\tilde C_{\rm back}$, $\tilde g_{\!f01}=\tilde U-\tilde C_{\rm back}$,
$\tilde g_{t01}=\tilde U-\tilde C_{\rm back}$ and $\tilde g_{b01}=\tilde U
+\tilde C_{\rm back}$, where $\tilde C_{\rm back}=C_{\rm back} a/(\pi v_f)$ is
the corresponding interaction factor.

When we include instead a forward interchain interaction, of parameter
$\tilde C_{\rm for}=C_{\rm for} a/(\pi v_f)$, the modulation of the bare
parameters writes $\tilde g_{02}=\tilde U+\tilde C_{\rm for}$,
$\tilde g_{\!f02}=\tilde U+\tilde C_{\rm for}$,
$\tilde g_{t02}=\tilde U-\tilde C_{\rm for}$ and
$\tilde g_{b02}=\tilde U-\tilde C_{\rm for}$.

Eventually, if we include both interaction, we only need to add the two
modulations.

\subsection{External excitation fields}

We note $\cal Z$ the three-legged couplings to external excitation fields.  We
will write ${\cal Z}^{\rm DW}_\alpha$, ($\alpha=C,S$) for Charge or Spin density
waves, and ${\cal Z}^{\rm SC(\Gamma)}_\alpha$, ($\alpha=s,t$) for singlet or
triplet superconducting states of symmetry $\Gamma=s,p,d,f,g$.  Again, we will
omit index $\alpha$ as soon as it is not necessary, and distinguish $z^{\rm
DW}_0$, which corresponds to $L^\dag_0R_0$ process, $z^{\rm DW}_\pi$
($\leftrightarrow L^\dag_\pi R_\pi$), $z^{\rm DW}_+$ ($\leftrightarrow
L^\dag_\pi R_0$) and $z^{\rm DW}_-$ ($\leftrightarrow L^\dag_0R_\pi$).  The
first process corresponds to an intraband mapping that relates 0-band states
(horizontally in the 2-$d$ representation of Fig.~\ref{para2d}), idem for the
second one with $\pi$-band ones, while the last ones are interband mappings
(biased in Fig.~\ref{para2d}); the same applies to ${\cal Z}^{\rm SC}$, except
that the processes now write $L_0R_0$, $L_\pi R_\pi$, $L_0R_\pi$ and $L_\pi
R_0$.  The definitions of all these couplings are detailed in%
~\ref{defcouplingsbis} and Fig.~\ref{figcouplingsbis}; from symmetry
considerations, one only needs $z_0$, $z_\pi$ and $z_+$; again, in the very case
of a ladder, we will see that $z_\pi$ can also be deduced from $z_0$, so we only
deal with two couplings per instability.

With our specific choice, the initial values of the couplings to external fields
are all $z_i=1$.

After this brief presentation of the model, we will now expound  the
classification of the different instabilities, according to their symmetries,
which we will study afterwards.

\section{Response functions}

To each external excitation field corresponds a susceptibility, which is the
response function of a pair operator. The corresponding order
parameter is the mean value of this operator. In order to classify the
different instabilities, one just needs to classify the pair operators.
Their symmetries are detailed in appendix~\ref{symmetries}.

We will first begin with SC instabilities.

\subsection{Superconducting instabilities}

\subsubsection{SC Hamiltonian}

Let us define the superconducting order parameters
$\Delta^{(\Gamma)}_\alpha({\bf X})=
\langle O^{(\Gamma)}_\alpha({\bf X})\rangle$, where the electron-electron pair
operator writes $O^{(\Gamma)}_\alpha=\sum_{\bf X'\sigma\sigma'}
\psi_{{\bf X},\sigma}\Gamma({\bf X},{\bf X'})
\psi_{{\bf X'},\sigma'}\tau^\alpha_{\sigma\sigma'}$,
with $\tau^s=\ii\sigma_y$ for singlet states,
$\tau^{t_x}=\ii\sigma_x\sigma_y=-\sigma_z$, $\tau^{t_y}=\ii\sigma_y\sigma_y=\ii I$
and $\tau^{t_z}=\ii\sigma_z\sigma_y=\sigma_x$ for triplet states ($\sigma_i$ are
the Pauli matrix, $I$ is the $2\times2$ identity matrix and $\ii$ is the
imaginary number). $\psi_{{\bf X},\sigma}$ is a real space wave function; since
electrons occupy discrete positions ${\bf X}=(ia,bj/2)$ ($i=1\cdot\cdot N$,
$j=\pm1$), we will rather write $\psi_{ij\sigma}$. The corresponding Hamiltonian
writes, in reciprocal space variables,
\begin{eqnarray}
H_{\rm SC}&=&\sum_{\scriptscriptstyle{\bf P}_1,{\bf P}_2\atop\sigma_1,\sigma_2}
\bar{\cal Z}^{\rm SC}({\scriptscriptstyle {\bf P}_1,{\bf P}_2,{\bf Q}})
\tau^{(\alpha)}\phi^{\phantom{\dag}}_{\bf Q}
\Psi^{\phantom{\dag}}_{{\bf P}_1\sigma_1}
\Psi^{\phantom{\dag}}_{{\bf P}_2\sigma_2}\nonumber\\
&+&{\cal Z}^{\rm SC}({\scriptscriptstyle {\bf P}_1,{\bf P}_2,{\bf Q}})
\bar\tau^{(\alpha)}\phi^\dag_{\bf Q}
\Psi^\dag_{{\bf P}_2\sigma_2}\Psi^\dag_{{\bf P}_1\sigma_1}
\label{HamiltonSC}
\end{eqnarray}
where ${\bf Q}=(Q_\Vert,Q_\perp)={\bf P}_1+{\bf P}_2$ is the interaction
momentum.

In order to simplify our expressions in this section, we write
$L_{p,\theta,\sigma}=L_{\theta,\sigma}(p-k_{f\theta})$
and $R_{p,\theta,\sigma}=L_{\theta,\sigma}(p+k_{f\theta})$,
where $\theta=0,\pi$. Notation $\int_L$ stands for the half band
integration $\int_{k_{f0}-{\pi\over a}}^{k_{f0}}$ or
$\int_{k_{f\pi}-{\pi\over a}}^{k_{f\pi}}$, which case we won't need to precise,
since what takes place at the band limit is not physically relevant, in this
system.

To each order parameter $\Delta({\bf X})$ corresponds an infinite number of
Fourier componants, depending on the reciprocal space variable $\bf Q$. We will
only keep componants ${\bf Q}=(0,0)$ and
${\bf Q}=(\pm\Delta k_f,{\pi\over b})$, so that $\epsilon({\bf k})$ and
$\epsilon({\bf Q-k})$ both lie in the physical band, close to the Fermi points.
We will write, in short, $O(0)$ and $O(\pi_\pm)$ the corresponding pair
operators.

Let us now classify the different operators, according to their symmetry, by
choosing the adequate $\Gamma$. The principles of the calculation and some
details are given in appendix~\ref{Fourier}.

\subsubsection{Singlet $s$ condensates}

The local pairing $\Gamma({\bf X},{\bf X'})=\delta_{ii'}\delta_{jj'}$
gives singlet condensates of $s$ symmetry. The pair operator reduces to
$$
O^{(s)}_s({\bf X})=2\psi_{\bf X\uparrow}\psi_{\bf X\downarrow}
=2\psi_{ij\uparrow}\psi_{ij\downarrow}\ .
 $$

$O(0)$ component corresponds to an intraband pairing, named 0-condensate
($\bf Q=0$, corresponding to $z^{\rm SC}_0$, see~\ref{defcouplingsbis})
and writes
\begin{eqnarray*}
O_s^{(s)}(0)&=&2\sum_i\psi_{i,1\uparrow}\psi_{i,1\downarrow}
+\psi_{i,-1\uparrow}\psi_{i,-1\downarrow}\\
&=&\int_L{adp\over2\pi} L_{p,0\uparrow}R_{-p,0\downarrow}
+L_{p,\pi\uparrow}R_{-p,\pi\downarrow}\\
&&\qquad+R_{-p,0\uparrow}L_{p,0\downarrow}
+R_{-p,\pi\uparrow}L_{p,\pi\downarrow}\ .
\end{eqnarray*}

$O(\pi_\pm)$ component corresponds to an interband pairing, named
$\pi$-condensate (${\bf Q}=(\pm\Delta k_f,{\pi\over b})$, corresponding to
$z^{\rm SC}_+$, see~\ref{defcouplingsbis}) and writes
\begin{eqnarray*}
&&\!\!\!\!\!\!\!\!O_s^{(s)}(\pi_\pm)=-2\ii\sum_i\e^{\mp\ii\Delta k_f ia}
\big(\psi_{i,1\uparrow}\psi_{i,1\downarrow}
-\psi_{i,-1\uparrow}\psi_{i,-1\downarrow}\big)\\
&&\!\!\!\!\!\!\!\!=-\ii\int_L{adp\over2\pi}
L_{p,0\uparrow}R_{\Delta k_f(1\pm1)-p,\pi\downarrow}
+L_{p,\pi\uparrow}R_{\Delta k_f(-1\pm1)-p,0\downarrow}\\
&&\qquad+R_{\Delta k_f(-1\pm1)-p,0\uparrow}L_{p,\pi\downarrow}
+R_{\Delta k_f(1\pm1)-p,\pi\uparrow}L_{p,0\downarrow}\ ;
\end{eqnarray*}
it is however antisymmetric with parity ($PO_s^{(s)}({\pi\over b})=
-O_s^{(s)}({\pi\over b})$); this comes from the $\e^{-\ii Q_\perp j b/2}$ factor
in the Fourier calculation, see details in~\ref{Fourier}.

The $s$ condensates are local in real space, see Fig.~\ref{spairing} (a).

If $\Gamma$ is replaced by $\delta_{i,i'\mp m}\delta_{jj'}$, one gets extended
$s$ states (in reciprocal space variables, the componants are multiplied by
$\cos(mPa)$ or some similar factor, see some examples in \ref{Fourier}).
However, we did not include these in our calculations.

\subsubsection{Singlet $d$ and $g$ condensates}

There are also two singlet condensates of $d$ and $g$ symmetry.

With $\Gamma=\delta_{i,i'}\delta_{j,-j'}$ (interchain pairing, with equal
positions on each chain), one gets another pair operator.
$O_s(\pi)$ component is zero for singlet condensate, while $O(0)$ component
corresponds to an intraband pairing (0-condensate) of $d$ symmetry, and writes
\begin{eqnarray*}
O_s^{(d)}(0)&=&2\sum_i\psi_{i,1\uparrow}\psi_{i,-1\downarrow}
+\psi_{i,-1\uparrow}\psi_{i,1\downarrow}\\
&=&\int_L{adp\over2\pi}
L_{p,0\uparrow}R_{-p,0\downarrow}
-L_{p,\pi\uparrow}R_{-p,\pi\downarrow}\\
&&\qquad+R_{-p,0\uparrow}L_{p,0\downarrow}
-R_{-p,\pi\uparrow}L_{p,\pi\downarrow}\ .
\end{eqnarray*}

With $\Gamma=\delta_{i,i'\mp1}\delta_{j,-j'}$, one gets a more complicated
pair operator. $O(\pi)$ component corresponds to an interband pairing
($\pi$-condensate) of $g$ symmetry, and writes
\begin{eqnarray*}\!\!\!\!\!\!\!\!
O_s^{(g)}(\pi_\pm)&=&-2\ii\sum_i\e^{\mp\ii\Delta k_f ia}
\big(\psi_{i,1\uparrow}\psi_{i+1,-1\downarrow}
-\psi_{i,-1\uparrow}\psi_{i+1,1\downarrow}\big)\\
&=&-\e^{\pm\ii{\Delta k_{\!f}a\over2}}
\int_L{adp\over2\pi}\sin(a(p-k_{f0}\mp{\Delta k_f\over2}))\\
\lefteqn{\big( L_{p-\Delta k_f,\pi\uparrow}R_{\pm\Delta k_f-p,0\downarrow}
+R_{\pm\Delta k_f-p,0\uparrow}L_{p-\Delta k_f,\pi\downarrow}}\\
\lefteqn{-L_{p,0\uparrow}R_{\Delta k_f(1\pm1)-p,\pi\downarrow}
-R_{\Delta k_f(1\pm1)-p,\pi\uparrow}L_{p,0\downarrow}\big)\ ;}
\end{eqnarray*}
be careful that the symmetry of the 0-condensate is $d_{x^2-y^2}$ (i.e. it
changes sign with $\tilde C$, see the definition afterwards), while that of the

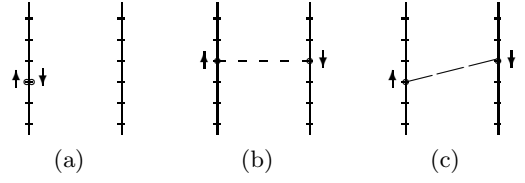
\begin{figure}[H]
\begin{center}
\begin{picture}(50,60)
\put(5,10){\line(0,1){50}}
\put(40,10){\line(0,1){50}}
\put(0,28){\vector(0,1){6.5}}
\put(10,35){\vector(0,-1){6.5}}
\multiput(3.5,14)(0,8){6}{\line(1,0){2.5}}
\multiput(38.5,14)(0,8){6}{\line(1,0){2.5}}
\put(4,30){\circle{2}}

\put(6,30){\circle{2}}
\put(20,0){\makebox(0,0){(a)}}
\end{picture}
\hspace{15pt}
\begin{picture}(50,60)
\put(5,10){\line(0,1){50}}
\put(40,10){\line(0,1){50}}
\put(0,35){\vector(0,1){6.5}}
\put(45,42){\vector(0,-1){6.5}}
\multiput(3.5,14)(0,8){6}{\line(1,0){2.5}}
\multiput(38.5,14)(0,8){6}{\line(1,0){2.5}}
\multiput(5,38)(8,0){5}{\line(1,0){3}}
\put(5,38){\circle{2}}
\put(40,38){\circle{2}}
\put(20,0){\makebox(0,0){(b)}}
\end{picture}
\hspace{15pt}
\begin{picture}(50,60)
\put(5,10){\line(0,1){50}}
\put(40,10){\line(0,1){50}}
\put(0,28){\vector(0,1){6.5}}
\put(45,42){\vector(0,-1){6.5}}
\multiput(3.5,14)(0,8){6}{\line(1,0){2.5}}
\multiput(38.5,14)(0,8){6}{\line(1,0){2.5}}
\multiput(5,30)(12,3){3}{\line(4,1){10}}
\put(5,30){\circle{2}}
\put(40,38){\circle{2}}
\put(20,0){\makebox(0,0){(c)}}
\end{picture}
\end{center}
\caption{Real space representation of SC condensate of singlet $s$ symmetry~(a),
$d$~(b) and $g$~(c)}
\label{spairing}
\end{figure}

$\pi$-condensate is both $d_{xy}$ (i.e. it changes sign with $p_x$ {\bf and}
$p_y$) and $d_{x^2-y^2}$; moreover, $d_{x^2-y^2}$ is imperfect on the
$\pi$-condensate (for instance, it maps a factor
sin($a(p-k_{f0}+{\Delta k_f\over2})$) onto
sin($a(-p-k_{f0}+{\Delta k_f\over2})$), which slightly differs); however, the
signs change according to $g$ symmetry. A real space representation of the
different condensate of singlet symmetry is given in Fig.~\ref{spairing}.

If the componants, in reciprocal variables, are multiplied by $\cos(mPa)$ (or
some similar factor), one gets extended $d$ condensates (this corresponds
to the harmonic classification).

\subsubsection{Triplet condensates}

One also finds triplet instabilities.

$\Gamma=\delta_{i,i'\mp1}\delta_{jj'}$ corresponds to the $p_x$ symmetry;
$O_t^{(p_x)}(0)$ corresponds to an intraband pairing (0-condensate), and
writes
\begin{eqnarray*}
O_t^{(p_x)}(0)&=&
\sum_{i\atop\sigma\sigma'}\big(\psi_{i,1\sigma}\psi_{i+1,1\sigma'}
+\psi_{i,-1\sigma}\psi_{i+1,-1\sigma'}\big)\tau^\alpha_{\sigma\sigma'}\\
&=&-\ii\int_L{adp\over2\pi}
\sum_{\sigma\sigma'}\big(L_{p,0\sigma}R_{-p,0\sigma'}\\
&&+L_{p-\Delta k_f,\pi\sigma}R_{\Delta k_f-p,\pi\sigma'}\big)\sin(a(p-k_{f0}))
\tau^\alpha_{\sigma\sigma'}\ .
\end{eqnarray*}

$O_t^{(p_x)}(\pi_\pm)$ corresponds to an interband pairing ($\pi$-condensate),
and writes
\begin{eqnarray*}
&&\quad O_t^{(p_x)}(\pi_\pm)=\\
&&-\ii\sum_{i\atop\sigma\sigma'}\e^{\mp\ii\Delta k_f ia}
\big(\psi_{i,1\sigma}\psi_{i+1,1\sigma'}
-\psi_{i,-1\sigma}\psi_{i+1,-1\sigma'}\big)\tau^\alpha_{\sigma\sigma'}\\
&&\qquad=-\e^{\pm\ii{\Delta k_{\!f} a\over2}}\int_L{adp\over2\pi}
\sum_{\sigma\sigma'}\big(L_{p,0\sigma}R_{\Delta k_f(1\pm1)-p,\pi\sigma'}\\
&&+L_{p-\Delta k_f,\pi\sigma}R_{\pm\Delta k_f-p,0\sigma'}\big)
\sin(a(p-k_{f0}\mp{\Delta k_f\over2}))\tau^\alpha_{\sigma\sigma'}\ ;
\end{eqnarray*}
be careful that, because of the factor $\e^{-\ii Q_\perp j b/2}$ in the Fourier
transform, this condensate is invariant under $P$.

With $\Gamma=\delta_{i,i'\mp1}\delta_{j,-j'}$, one gets an intraband pairing
(0-condensate) of symmetry $f_x$, given by the $O_t^{(f_x)}(0)$ component
\begin{eqnarray*}
O_t^{(f_x)}(0)&=&\sum_{i\atop\sigma\sigma'}
\big(\psi_{i,1\sigma}\psi_{i+1,-1\sigma'}
+\psi_{i,-1\sigma}\psi_{i+1,1\sigma'}\big)\tau^\alpha_{\sigma\sigma'}\\
&=&\ii\int_L{adp\over2\pi}
\sum_{\sigma\sigma'}\big(L_{p-\Delta k_f,\pi\sigma}R_{\Delta k_f-p,\pi\sigma'}\\
&&\qquad-L_{p,0\sigma}R_{-p,0\sigma'}\big)\sin(a(p-k_{f0}))
\tau^\alpha_{\sigma\sigma'}\ ;
\end{eqnarray*}
note that $d_{x^2-y^2}$ is again imperfect.

\begin{figure}[H]
\begin{center}
\begin{picture}(50,60)
\put(5,10){\line(0,1){50}}
\put(40,10){\line(0,1){50}}
\put(35,28){\vector(0,1){6.5}}
\put(45,35){\vector(0,1){6.5}}
\multiput(3.5,14)(0,8){6}{\line(1,0){2.5}}
\multiput(38.5,14)(0,8){6}{\line(1,0){2.5}}
\put(40,30){\circle{2}}
\put(40,38){\circle{2}}
\put(20,0){\makebox(0,0){(a)}}
\thicklines
\put(40,30){\line(0,1){8}}
\end{picture}
\hspace{15pt}
\begin{picture}(50,60)
\put(5,10){\line(0,1){50}}
\put(40,10){\line(0,1){50}}
\put(0,28){\vector(0,1){6.5}}
\put(45,35){\vector(0,1){6.5}}
\multiput(5,30)(12,3){3}{\line(4,1){10}}
\multiput(3.5,14)(0,8){6}{\line(1,0){2.5}}
\multiput(38.5,14)(0,8){6}{\line(1,0){2.5}}
\put(5,30){\circle{2}}
\put(40,38){\circle{2}}
\put(20,0){\makebox(0,0){(b)}}
\end{picture}
\hspace{15pt}
\begin{picture}(50,60)
\put(5,10){\line(0,1){50}}
\put(40,10){\line(0,1){50}}
\put(0,28){\vector(0,1){6.5}}
\put(45,28){\vector(0,1){6.5}}
\multiput(3.5,14)(0,8){6}{\line(1,0){2.5}}
\multiput(38.5,14)(0,8){6}{\line(1,0){2.5}}
\multiput(5,30)(8,0){5}{\line(1,0){3}}
\put(5,30){\circle{2}}
\put(40,30){\circle{2}}
\put(20,0){\makebox(0,0){(c)}}
\end{picture}
\end{center}
\caption{Real space representation of SC condensate of triplet symmetry
$p_x$~(a), $f_x$~(b) and $f_y$~(c)}
\label{tpairing}
\end{figure}
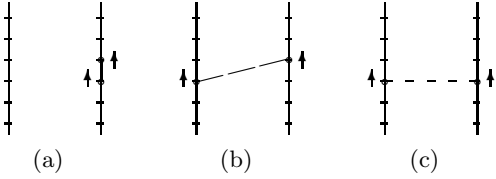

With $\Gamma=\delta_{i,i'}\delta_{j,-j'}$, one gets an interband pairing
($\pi$-condensate) of symmetry $f_y$, given by the $O_t^{(f_y)}(\pi_\pm)$
component
\begin{eqnarray*}
&&\!\!\!\!\!\!\!\!\!\!\!\!\!\!O_t^{(f_y)}(\pi_\pm)=
-\ii\sum_{i\atop\sigma\sigma'}\e^{\mp\ii\Delta k_f ia}
\big(\psi_{i,1\sigma}\psi_{i,-1\sigma'}
-\psi_{i,-1\sigma}\psi_{i,1\sigma'}\big)\tau_{\sigma\sigma'}\\
&&\qquad\ =\;\ii\int_L{adp\over2\pi}
\sum_{\sigma\sigma'}\big(L_{p,0\sigma}R_{\Delta k_f(1\pm1)-p,\pi\sigma'}\\
&&\qquad\qquad\qquad
-L_{p,\pi\sigma}R_{\Delta k_f(-1\pm1)-p,0\sigma'}\big)\tau_{\sigma\sigma'}\ ;
\end{eqnarray*}
note that $p_y$ antisymmetry is an internal one and does not account on the
exponential factor, in the Fourier transform. A real space representation of
the different condensate of triplet symmetry is given in Fig.~\ref{tpairing}.

Extended states of the same symmetries can be obtained exactly the same
way as for singlet superconducting operators.

\subsection{Density wave instabilities}

\subsubsection{DW Hamiltonian}

We have also investigated site density waves instabilities, defined by the order
parameter $\Delta^{\rm DW}_{\rm site}({\bf X})=$
$\sum_{\sigma\sigma'}
\langle\psi^\dag_{\bf X,\sigma}\psi^{\phantom{\dag}}_{\bf X,\sigma'}\rangle
\tau^\alpha_{\sigma\sigma'}$, with $\tau^C=I$ for CDW, and
$\tau^{S_x}=\sigma_x$, $\tau^{S_y}=\sigma_y$ and $\tau^{S_z}=\sigma_z$,
for SDW, as well as bond density waves instabilities, defined by the order
parameter $\Delta^{\rm DW}_{\rm bond}({\bf X})=
\sum_{\sigma\sigma'}\langle\psi^\dag_{\bf X,\sigma}
\psi^{\phantom{\dag}}_{\bf X+1_\Vert,\sigma'}\rangle
\tau^\alpha_{\sigma\sigma'}$, where ${\bf 1}_\Vert=(1,0)$. These couplings
are intrachain, we distinguish intraband and interband ones. We also
investigated interchain couplings, defined by the order parameters
$\Delta^{\rm DW}_{\rm site}({\bf X})=
\sum_{\sigma\sigma'}\langle\psi^\dag_{\bf X,\sigma}
\psi^{\phantom{\dag}}_{\bf X+G,\sigma'}\rangle\tau^\alpha_{\sigma\sigma'}$
and $\Delta^{\rm DW}_{\rm bond}({\bf X})=\sum_{\sigma\sigma'}
\langle\psi^\dag_{{\bf X},\sigma}
\psi^{\phantom{\dag}}_{\bf X+G',\sigma'}\rangle
\tau^\alpha_{\sigma\sigma'}$, where $G$ or
$G'\in\{\bf 1_\perp,1_\Vert+1_\perp\}$ and ${\bf 1}_\perp=(0,1)$.

The corresponding Hamiltonian writes, in reciprocal space variables,
\begin{eqnarray}
H_{\rm DW}&=&\sum_{\scriptscriptstyle{\bf P}_1,{\bf P}_2\atop\sigma_1,\sigma_2}
\bar{\cal Z}^{\rm DW}({\scriptscriptstyle {\bf P}_1,{\bf P}_2,\bf Q})
\tau^{(\alpha)}
\phi^\dag_{\bf Q}\Psi^\dag_{{\bf P}_1\sigma_1}
\Psi^{\phantom{\dag}}_{{\bf P}_2\sigma_2}\nonumber\\
&+&{\cal Z}^{\rm DW}({\scriptscriptstyle {\bf P}_1,{\bf P}_2,\bf Q})
\bar\tau^{(\alpha)}\phi^{\phantom{\dag}}_{\bf Q}\Psi^\dag_{{\bf P}_2\sigma_2}
\Psi^{\phantom{\dag}}_{{\bf P}_1\sigma_1}\ .
\label{HamiltonDW}
\end{eqnarray}
The construction of the response functions for these instabilities is very
similar to that of the superconducting instabilities. So, we will only give
the Fourier componants of the electron-hole pair operator for
${\bf Q}=(-2k_{f0},0)$, ${\bf Q}=(-2k_{f\pi},0)$
and ${\bf Q}=(-k_{f0}-k_{f\pi},{\pi\over b})$.

SDW and CDW operators only differ by the spin factor (matrix $\sigma$ or $I$),
so we will also omit this factor.

\subsubsection{DW response function}

The intraband response functions write then
\begin{eqnarray*}
&&\!\!\!\!\!\!\!\!O_{\rm site}(-2k_{f0},0)=\int_L{adp\over2\pi}
R^\dag_{p,0\sigma}L^{\phantom{\dag}}_{p,0\sigma'}
+R^\dag_{p+\Delta k_f,\pi\sigma}L^{\phantom{\dag}}_{p-\Delta k_f,\pi\sigma'}\\
&&\!\!\!\!\!\!\!\!\!\!\!
O_{\rm bond}(-2k_{f0},0)=\ii\int_L{adp\over2\pi}\sin(ap)\e^{-\ii k_{f0}a}\\
&&\qquad\qquad\qquad(R^\dag_{p,0\sigma}L^{\phantom{\dag}}_{p,0\sigma'}
+R^\dag_{p+\Delta k_f,\pi\sigma}L^{\phantom{\dag}}_{p-\Delta k_f,\pi\sigma'})
\end{eqnarray*}
and
\begin{eqnarray*}
&&\!\!\!\!\!\!\!\!O_{\rm site}(-2k_{f\pi},0)=\int_L{adp\over2\pi}
R^\dag_{p-\Delta k_f,0\sigma}L^{\phantom{\dag}}_{p+\Delta k_f,0\sigma'}
+R^\dag_{p,\pi\sigma}L^{\phantom{\dag}}_{p,\pi\sigma'}\\
&&\!\!\!\!\!\!\!\!\!\!\!
O_{\rm bond}(-2k_{f\pi},0)=\ii\int_L{adp\over2\pi}\sin(ap)\e^{-\ii k_{f\pi}a}\\
&&\qquad\qquad\qquad
(R^\dag_{p-\Delta k_f,0\sigma}L^{\phantom{\dag}}_{p+\Delta k_f,0\sigma'}
+R^\dag_{p,\pi\sigma}L^{\phantom{\dag}}_{p,\pi\sigma'})\ .
\end{eqnarray*}

The interband response function writes
\begin{eqnarray*}
&&O_{\rm site}(-k_{f0}-k_{f\pi},{\pi\over b})=\ii\int_L{adp\over2\pi}
R^\dag_{p,0\sigma}L^{\phantom{\dag}}_{p,\pi\sigma'}+
R^\dag_{p,\pi\sigma}L^{\phantom{\dag}}_{p,0\sigma'}\\
&&\!\!\!O_{\rm bond}(-k_{f0}-k_{f\pi},{\pi\over b})=-\int_L{adp\over2\pi}
\e^{a{k_{f0}+k_{f\pi}\over2}}\\
&&\quad\sin(a(p-{\Delta k_f\over2}))
(R^\dag_{p,\pi\sigma}L^{\phantom{\dag}}_{p,0\sigma'}+
R^\dag_{p+\Delta k_f,0\sigma}L^{\phantom{\dag}}_{p+\Delta k_f,\pi\sigma'})
\ .
\end{eqnarray*}

The above response function are intrachain ones. The way we have written them
one just needs to add a minus sign before the first (or second) term of
all intrachain operators, to obtain all interchain ones.

\section{Renormalization Group equations}

\subsection{Choice of the RG scheme}

We have used the One Particle Irreducible (OPI) scheme (Ref.~\cite{Honerkamp,
Honerkamp2,Binz}), to calculate all diagrams, in a one-loop expansion.  We use
the flow parameter $\Lambda=\Lambda_o\e^{-\ell}$. We do not renormalize $v_f$
nor $k_{f0}$ or $k_{f\pi}$.

\subsection{$k_\Vert$ dependent equations}

\subsubsection{$k_\Vert$ dependence of the couplings}
\label{dependance}

In this system, the interband backward scattering $g_b$ plays a particular role.
Due to momentum conservation, it is not possible to put all its arguments onto
the Fermi points.  This indicates that $g_b$ is not a low energy process.
However, it {\em intervenes} in the renormalization of low energy processes.
For instance, in the renormalization of $g_0$, $dg_0\over d\ell$ gives a
contribution containing a $g_{b0}$ scattering, with a factor
$\Lambda\over\Lambda+v_f\Delta k_f$.  This contribution is exponentially
suppressed, as $\Lambda\ll v_f\Delta k_f=2t_\perp$.  It can thus be neglected
as soon as $v_f\Delta k_f$ is of the order or bigger than the initial bandwidth
$2\Lambda_0$.  On the other hand, as shown by Fabrizio\cite{Fabrizio}, the $g_b$
process has to be taken into account if $v_f\Delta k_f$ is much smaller than
$\Lambda_0$.

Thereupon, in order to calculate the renormalization of $g_{b0}$ properly,
couplings $g_0$, $g_f$ or $g_t$ with specific $k_\Vert$ dependence are
needed. For instance, $dg_{b0}\over d\ell$ gives a Peierls diagram~:
\begin{center}
\begin{picture}(80,80)
\qbezier(25,60)(70,40)(25,20)
\qbezier(50,60)(10,40)(50,20)
\put(33,56){\vector(2,-1){0}}
\put(30,42){\vector(0,1){0}}
\put(31,23){\vector(-2,-1){0}}
\put(47,58){\vector(2,1){0}}
\put(48,38){\vector(0,-1){0}}
\put(43,24){\vector(-2,1){0}}
\put(20,66){\makebox(0,0){$\scriptscriptstyle k_{f0}$}}
\put(15,45){\makebox(0,0){$\scriptscriptstyle {\Lambda\over v_f}-k_{f0}$}}
\put(15,35){\makebox(0,0){$\scriptscriptstyle +2\Delta k_f$}}
\put(20,15){\makebox(0,0){$\scriptscriptstyle k_{f\pi}$}}
\put(66,66){\makebox(0,0){$\scriptscriptstyle -k_{f0}+2\Delta k_f$}}
\put(66,40){\makebox(0,0){$\scriptscriptstyle {\Lambda\over v_f}+k_{f0}$}}
\put(57,15){\makebox(0,0){$\scriptscriptstyle -k_{f\pi}$}}
\put(38,60){\makebox(0,0){$\scriptscriptstyle g_0$}}
\put(38,19){\makebox(0,0){$\scriptscriptstyle g_{b0}$}}
\end{picture}
\end{center}
including coupling $g_0({\scriptstyle k_{f0},-k_{f0}+2\Delta k_f,
-k_{f0}+2\Delta k_f,k_{f0}})$, with arguments that remain separated from the
Fermi points, even in the limit $\Lambda\to0$, let us write it $g_{01}$. This
coupling separates from coupling
$g_0({\scriptstyle k_{f0},-k_{f0},-k_{f0},k_{f0}})$, with all arguments at the
Fermi points, which we will write $g_{00}$; therefore $k_\Vert$ dependence
is influential. This can be proved by comparing their renormalization,
in the Cooper channel. For instance, $dg_{00}\over d\ell$ gives, in the Cooper
channel, a term proportional to $g_0^2+g_t^2$ with a {\em constant} factor,
which is present all the way down to $\Lambda\to0$. On the contrary,
$dg_{01}\over d\ell$ gives, in the Cooper channel, a term with a factor
${2\Lambda\over2\Lambda+v_f |P_1+P_2|}={\Lambda\over\Lambda+v_f\Delta k_f}$;
the renormalization of $g_{01}$ in the Cooper channel is thus exponentially
suppressed, when $\Lambda\to0$ (for $g_{00}$, the total incoming momentum is
$P_1+P_2=0$).

We have proved that different $g_0$ couplings separate during the flow, so the
$k_\Vert$ dependence is hence capable to have an effect
during the flow, when it is taken into account. This generalizes for $g_f$,
$g_t$ and even $g_b$ couplings.

All this differs completely from the one-dimensional case, where the
renormalization of the coupling with all momenta on the Fermi points
is only governed by processes with momenta $\pm k_f\pm{\Lambda\over v_f}$, which
always fall on the Fermi points when $\Lambda\to0$.  In our case, it is not
possible to apply the same argument as soon as $t_\perp\ll\Lambda_o$. Indeed, we
will see, in the following, that one gets different results, depending on
whether we take the $k_\Vert$ dependence of the couplings into account or
not.

The $k_\Vert$ dependence can be observed, when $\Lambda$ is large
(and till $\Lambda>v_f\Delta k_f$), by the separation of the
different scattering couplings $g_0$, $g_{f0}$ and $g_{t0}$. On the contrary,
if one puts $g_b=0$ at $\ell=0$, this dependence is suppressed, and the
system becomes purely one-dimensional for small values of $t_\perp$ (in that
case, $g_b$ remains 0 for all $\Lambda$ and the RG equations simplify
drastically, although they differ from the one-dimension case).

\subsubsection{$k_\Vert$ representation of the couplings $g$}
\label{rules}

In order to write explicit $k_\Vert$-dependent RG equations, one needs to
define a consistent and  detailed $k_\Vert$ representation of the couplings
$g$.

Let us first note that ${\cal G}(P_1,P_2,P'_2,P'_1)$ corresponds to
$R^\dag(P_1)L^\dag(P_2)L(P'_2)R(P'_1)$, where $P_i$ are the absolute momenta in
the $\Vert$ direction. We then define the relative momenta
$p_1=P_1-k_{f\theta_1}$, $p_2=P_2+k_{f\theta_2}$, $p_3=P_3+k_{f\theta_3}$ and
$p_4=P_4-k_{f\theta_4}$, and write, correspondingly, $g(p_1,p_2,p'_2,p'_1)$.
We also introduce variables $c=p_1+p_2=p'_1+p'_2+d$, $l=p_1-p'_1=p'_2-p_2+d$
and $p=p_1-p'_2=p'_1-p_2+d$, where $d=-2\Delta k_f$ for $g_{b0}$,
$d=2\Delta k_f$ for $g_{b\pi}$, and $d$ is zero otherwise, and then
write, correspondingly, $g(c,l,p)$ ($d$ is implicit).

At the beginning of this section, we have found in a diagram a particular
coupling\penalty-50000
$g_0({\Lambda\over v_f},2\Delta k_f,0,2\Delta k_f+{\Lambda\over v_f})$.
When $\Lambda\to0$, we get $g_0(0,2\Delta k_f,0,2\Delta k_f)$ (which also writes
$g_0(2\Delta k_f,-2\Delta k_f,0)$ in $(c,l,p)$ notation).  Note that some
arguments are shifted by $\pm 2\Delta k_f$, compared to the coupling
$g_0(0,0,0,0)$ with all momenta on the Fermi points.

This could be easily generalized for all couplings $g$. So, in order to get a
reasonable number of couplings, we have done the following approximation~: all
terms $\pm{\Lambda\over v_f}$, in all diagrams, are replaced by their
$2\Delta k_f$ part (i.e. by $2\Delta k_f\left[{\Lambda\over 2\Delta k_f
v_f}\right]$, where $[x]$ is the biggest integer $\le x$). Then, it
follows that we only get couplings $g(p_1,p_2,p'_2,p'_1)$ (or $g(c,l,p)$), where
all variables $p_i$ (or $c,l,p$) are multiples of $2\Delta k_f$.

\subsubsection{$k_\Vert$ representation of the couplings $z$}

All the preceding procedure generalizes to the couplings $z$ as well.

We first introduce a $(k,c,p)$ representation, similarly, with $c=p_1+p_2$,
$p=p_2-p_1$ and $k=p_1$, where $p_1$ and $p_2$ are defined on
Fig.~\ref{figcouplingsbis} and write, correspondingly, $z^{\rm SC}(c,k)$ or
$z^{\rm DW}(p,k)$.

Then, we apply the same approximation in order to get couplings, where all
variables $(k,c,p)$ are multiples of $2\Delta k_f$.

The same conclusion applies to these couplings, proving that their
$k_\Vert$ dependence is also influential.

\subsubsection{RG equations}

Finally, we calculate the RG flow of the separated following couplings~:
$g_0(0,0,0,0)$,\penalty-50000
$g_0(-2\Delta k_f,0,0,-2\Delta k_f)$, $g_0(0,2\Delta k_f,0,2\Delta k_f)$, etc.
as well as $z(0,0)$, $z(2\Delta k_f,0)$, etc.

The exact RG equations, including all $\Vert$ components, are given in
\ref{eqRGg}, for the $g(c,l,p)$, and in \ref{eqRGz}, for the $z^{\rm SC}(c,k)$
and $z^{\rm DW}(p,k)$.

In order not to solve an infinite number of equations, we have reduced
the effective bandwidth of the renormalized couplings to
$4n_{\rm max}\Delta k_f$, where $n_{\rm max}$ is an integer, by projecting all
momenta lying out of the permitted band, back into it, according to a specific
truncation procedure that will be explained after.

We have performed our calculations with $n_{\rm max}=2$, 3 or 4, and the results
rapidly converge as $n_{\rm max}$ is increased.

\subsubsection{Susceptibility equations}

To each instability corresponds a susceptibility. We will write
$\chi\lower1.6pt\hbox{${}_\alpha^{\rm SC(\Gamma)}$}$ the different SC ones
and $\chi_\alpha^{\rm DW}$ the different SDW or CDW ones.

The susceptibilities have no $k_\Vert$ dependence. However, couplings $z$
with {\em different} $k_\Vert$ variables appear in their RG equations,
which we give in \ref{eqRGc}.

Referring to the transverse component of the interaction vectors, we will write
$\chi(0)$ the instabilities corresponding to intraband processes, and
$\chi({\pi\over b})$ those corresponding to interband ones.

We use several symmetries, to reduce the number of couplings. Because of the
$k_\Vert$ dependence, it is not as easy to apply them as in ordinary
cases. We give here some indications, which are completed in appendix.

\subsection{Symmetries}

\subsubsection{Ordinary symmetries}

We apply $C$ the conjugation symmetry ($C:{\bf r}\to {\bf r},
\;{\bf p}\to-{\bf p},\; \sigma\to-\sigma$), $A$ the (antisymmetrical) exchange
between incoming particles, $A'$ the exchange between outgoing particles,
$P$ the space parity ($P:{\bf r}\to-{\bf r},\; {\bf p}\to-{\bf p},\;
\sigma\to\sigma$) and $S$ the spin rotation ($S:\sigma\to-\sigma$). We will also
use $p_x$, $p_y$ (the mirror symmetries in the $\Vert$ and $\perp$
directions), $f_x$ and $f_y$. Note that $P=p_xp_y$.

$H_{\rm cin}$ (Eq.~(\ref{Hamiltoncin})) and $H_{\rm int}$
(Eq.~(\ref{Hamiltonint}))
satisfy all these symmetries, whereas SC instabilities, governed
by $H_{\rm SC}$ (Eq.~(\ref{HamiltonSC})), are not invariant under
$S$ or $C$, which allows a natural classification of the states, and DW
instabilities, governed by $H_{\rm DW}$ (Eq.~(\ref{HamiltonDW})),
do satisfy $CS$, $AS$ or $A'S$, but not $C$, $A$, $A'$ nor $S$ symmetries.

All the relations satisfied by $\cal G$ or $\cal Z$ couplings are detailed in
\ref{symbasic}. From what precedes, one will not be surprised that those for
$\cal G$ couplings are simpler and less sophisticated than those for $\cal Z$
ones.

\paragraph*{Relations of couplings $g$}

We are not interested here in the symmetries that relate, for instance, a $LRLR$
coupling to a $RLLR$ one. Instead, we only keep $RLLR$ couplings and deduce all
the symmetries that keep this order.

We then apply them to every coupling $g_0$, $g_\pi$, etc., and find, altogether,
exactly two independent relations for each one, which write, in $(c,l,p)$
notation,

\renewcommand{\theequation}{G-\arabic{equation}}
\setcounter{equation}{0}
\begin{equation}
\left.\begin{array}{rcl}
g_i(c,l,p)&=&g_i(-c,l,p)\quad i=0,\pi,t0,t\pi\\
g_{\!f\pi}(c,l,p)&=&g_{\!f0}(-c,l,p)\\
g_{b\pi}(c,l,p)&=&g_{b0}(-c,l-2\Delta k_f,p-2\Delta k_f)
\end{array}\right\}
\end{equation}
\begin{equation}
\left.\begin{array}{rcl}
g_i(c,l,p)&=&g_i(c,-l,p)\quad i=0,\pi,f\hspace{-.1em}0,f\hspace{-.1em}\pi\\
g_{t\pi}(c,l,p)&=&g_{t0}(c,-l,p)\\
g_{b\pi}(c,l,p)&=&g_{b0}(c-2\Delta k_f,-l,p-2\Delta k_f)
\end{array}\right\}
\end{equation}

One observes that G-1 relates $g_{\!f0}$ to $g_{\!f\pi}$ and $g_{b0}$ to
$g_{b\pi}$, while G-2 relates $g_{t0}$ to $g_{t\pi}$ and $g_{b0}$ to
$g_{b\pi}$. The combination of G-1 and G-2 thus relates $g_{\!f0}$ to $g_{\!f\pi}$
and $g_{t0}$ to $g_{t\pi}$.

\paragraph*{Relations of couplings $z$}

We similarly deduce all symmetries that keep the $LR$ order; we apply them to
every coupling $z_0$, $z_\pi$, etc., and find only one relation for each one,
which writes, in $(k,c,p)$ notation,

\renewcommand{\theequation}{Z-SC-\arabic{equation}}
\setcounter{equation}{0}
\begin{equation}
\left.\begin{array}{rcl}
z^{\rm SC}_{s\,\theta}(c,k)&=&z^{\rm SC}_{s\,\theta}(-c,k-c)\quad\theta=0,\pi\\
z^{\rm SC}_{t0}(c,k)&=&z^{\rm SC}_{t\pi}(-c,k-c)\\
z^{\rm SC}_{s-}(c,k)&=&\pm z^{\rm SC}_{s+}(-c,k-c)\\
z^{\rm SC}_{t-}(c,k)&=&\pm z^{\rm SC}_{t+}(-c,k-c)
\end{array}\right\}
\end{equation}
where $\pm$ reads + for $\Gamma=s$ or $\Gamma=p_x$ and
$-$ for $\Gamma=g$ or $\Gamma=f_y$. Note that
$z_0^{\rm SC}$ or $s_\pi^{\rm SC}$ correspond to intraband condensates, while
$z_+^{\rm SC}$ or $s_-^{\rm SC}$ correspond to interband ones.
\renewcommand{\theequation}{Z-DW-\arabic{equation}}
\setcounter{equation}{0}
\begin{equation}
\left.\begin{array}{rcl}
z^{\rm DW}_\theta(p,k)&=&\pm z^{\rm DW}_\theta(p,-p-c)\quad \theta=0,\pi\\
z^{\rm DW}_-(p,k)&=&\pm z^{\rm DW}_+(p,-p-c)
\end{array}\right\}
\end{equation}
where $\pm$ reads + for site ordering and $-$ for bond ordering.

One observes that Z-1 relates $z_+$ to $z_-$.

\subsubsection{Supplementary symmetry}

As we already noted, ordinary symmetries do not relate $g_0$ to $g_\pi$, nor
$z_0$ to $z_\pi$. However, since we choose identical bare values at $\ell=0$,
and since the RG equations are symmetrical, we observe an effective symmetry
between these couplings~: we will show here that this does not occur by chance,
but that it follows a specific symmetry $\tilde C$, which only applies to the
ladder system.

$\tilde C$ is a kind of conjugation~: it generalizes the electron-hole
symmetry as follows.

\begin{figure}[H]
\begin{picture}(80,80)
\put(0,40){\vector(1,0){80}}
\put(80,48){\makebox(0,0){$P$}}
\thicklines
\put(65,65){\line(-1,-1){50}}
\put(60,60){\circle*{3}}
\put(20,20){\circle*{3}}
\put(58,32){\makebox(0,0){$k_f+p$}}
\put(18,48){\makebox(0,0){$k_f-p$}}
\linethickness{0.2pt}
\put(60,38){\line(0,1){28}}
\put(20,14){\line(0,1){28}}
\put(40,0){(a)}
\end{picture}
\hspace{1pt}
\begin{picture}(125,80)
\put(35,40){\vector(1,0){85}}
\put(120,48){\makebox(0,0){$P$}}
\multiput(39,15)(12.3,10.5){5}{\line(6,5){10}}
\thicklines
\put(89,65){\line(-1,-1){50}}
\put(98,65){\line(-1,-1){50}}
\put(95,62){\circle*{3}}
\put(42,18){\circle*{3}}
\put(69,40){\circle*{3}}
\put(91,32){\makebox(0,0){$k_{f0}+p$}}
\put(37,48){\makebox(0,0){$k_{f\pi}-p$}}
\linethickness{0.2pt}
\put(95,38){\line(0,1){28}}
\put(42,14){\line(0,1){28}}
\put(80,0){(b)}
\end{picture}
\caption{Symmetry around the Fermi points~: (a)~in a 1-band
system; (b)~in a 2-band system}
\label{etrou}
\end{figure}
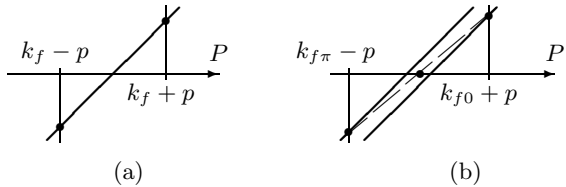

Let us first consider the case of a single band one-dimensional system; we
find an electron-hole symmetry, described in Fig.~\ref{etrou}~(a).

For $R$ particles, it conjugates an electron with momentum $k_f+p$ and a hole
with momentum $k_f-p$.  In the momentum space, it is a symmetry around $k_f$.
One can write $\tilde C\psi_p=\psi^\dag_{2k_f-p}$ and $\tilde
C\psi^\dag_p=\psi_{2k_f-p}$. For $L$ particles, the same relation applies if
one simply changes $k_f$ into $-k_f$.

Let us now return to the two-band system. This symmetry generalizes by turning
the momenta around the isobarycenter of the Fermi points, as shown in
$\Vert$ space in Fig.~\ref{etrou}~(b). For $R$ particles, $(k_{f0}+k_{f\pi})/2$
points to the isobarycenter and you now get
$\tilde C\psi_p=\psi^\dag_{k_{f0}+k_{f\pi}-p}$ etc.

In the two-dimensional representation, $\tilde C$-symmetry is a point
symmetry around $(\pm{k_{f0}+k_{f\pi}\over2},{\pi\over2b})$, as shown on
Fig.~\ref{para2d} (the sign depends on whether it is a $R$ or $L$ momentum).

Actually, there is an alternative symmetry, which we write $\notC$, which also
maps the band $k_\perp=0$ onto the band $\pi$~: it is a translation by the
vector $(\pm(k_{f\pi}-k_{f0}),{\pi\over b})$, as shown on Fig.~\ref{para2d}.
Some of the bare couplings satisfy this $\notC$ symmetry
($g_0$, $g_{\!f}$, $g_t$, $z_0$ or $z_\pi$), but some
don't ($g_b$, $z_+$, $z_-$). Since interactions are mixing all couplings
as soon as the flow parameter $\ell>0$, the renormalized couplings will break
the $\notC$ symmetry. Therefore, we can't use it.

On the contrary, one verifies that $H_{\rm cin}$, $H_{\rm int}$, $H_{\rm SC}$
and $H_{\rm DW}$ are
invariant under $\tilde C$. The induced relations satisfied by $\cal G$ or
$\cal Z$ couplings are detailed in \ref{symsupp}.

In fact, $\tilde C$ and $\notC$ weakly correspond to the $d_{x^2-y^2}$
symmetry in two dimensions.

\paragraph*{Supplementary relation of couplings $g$}

It is straightforward that $\tilde C$ keeps the $RLLR$ order when one applies
it to any coupling $g_i$; so we find a new relation for each
one, which writes, in $(c,l,p)$ notation,
\renewcommand{\theequation}{G-\arabic{equation}}
\setcounter{equation}{2}
\begin{equation}
\left.\begin{array}{rcl}
g_\pi(c,l,p)&=&g_0(-c,-l,-p)\\
g_{\!f\pi}(c,l,p)&=&g_{\!f0}(-c,-l,-p)\\
g_{t\pi}(c,l,p)&=&g_{t0}(-c,-l,-p)\\
g_{b\pi}(c,l,p)&=&g_{b0}(-c,-l,-p)
\end{array}\right\}
\end{equation}
One verifies that $g_0$ and $g_\pi$ are related; in fact, G-3 relates
all $0$-couplings to $\pi$-couplings.

\paragraph*{Supplementary relation of couplings $z$}

Similarly, $\tilde C$ keeps $LR$ order when we apply it to any coupling
$z_i$; so we find a new relation for each one, which writes, in
$(k,c,p)$ notations,
\renewcommand{\theequation}{Z-SC-\arabic{equation}}
\setcounter{equation}{1}
\begin{equation}
\left.\begin{array}{rcl}
z^{\rm SC}_\pi(c,k)&=&\pm z^{\rm SC}_0(-c,-k)\\
z^{\rm SC}_-(c,k)&=&\pm z^{\rm SC}_+(-c,-k)
\end{array}\right\}
\end{equation}
where $\pm$ reads + for $\Gamma=s$ or $\Gamma=p_x$ and
$-$ for $\Gamma=d,g$ or $\Gamma=f_x,f_y$.
\renewcommand{\theequation}{Z-DW-\arabic{equation}}
\setcounter{equation}{1}
\begin{equation}
\left.\begin{array}{rcl}
z^{\rm DW}_\pi(p,k)&=&\pm z^{\rm DW}_0(-p,-k)\\
z^{\rm DW}_-(p,k)&=&\pm z^{\rm DW}_+(-p,-k)
\end{array}\right\}
\end{equation}
where $\pm$ reads + for site ordering and $-$ for bond ordering.

Again, one verifies that $z_0$ and $z_\pi$ are related (as well as $z_+$ and
$z_-$).

\subsection{Truncation}

Understanding symmetry relations does not only help us to reduce drastically the
number of couplings, it is also an essential tool to make a proper truncation
procedure, as we will explain now.

\subsubsection{Triplet notation}

Let us first introduce a useful notation for the $k_\Vert$ dependence.

We have already defined the relative momentum representation
$g(p_1,p_2,p'_2,p'_1)$, as well as the $g(c,l,p)$ notation, and explained how
to keep only couplings, the arguments of which are all multiples of $2\Delta
k_f$. We will focus on the $(c,l,p)$ notation and write $c=2n_1\Delta k_f$,
$l=2n_2\Delta k_f$, $p=2n_3\Delta k_f$, with $(n_1,n_2,n_3)\in\Z^3$.

In short, we can write $g_i(n_1,n_2,n_3)$ ($i=0,f\hspace{-0.1em}0,t0,b0$), where
$(n_1,n_2,n_3)$ is called a triplet. Mind that, using symmetry relations, two
triplets $(n_1,n_2,n_3)$ and $(n'_1,n'_2,n'_3)$ can represent the same coupling.
One says that these triplets belong to the same symmetry orbit (or symmetry
class). Mind also that the orbits are different for each coupling $g_0$,
$g_{\!f0}$, $g_{t0}$ and $g_{b0}$.

It would take too long to give an exhausted list of these orbits. Let us just
observe that $(0,0,0)$'s orbit has only one element (itself), except for $g_b$,
the orbits of which we detailed in appendix~\ref{orbites}.

\subsubsection{Choice of the truncation procedure}

Obviously, one needs only renormalize one coupling per orbit. From the
fundamental rules, explained in \ref{rules}, it follows that, even if one
starts with only $g_0(0,0,0), g_{\!f0}(0,0,0)$ and $g_{t0}(0,0,0)$, the RG
equations will generate an infinite number of orbits. So, we will only keep
couplings which satisfy $|n_i|\le n_{\rm max}$ (for a given $n_{\rm max}$); but
even so, in the RG equations of some orbits intervene couplings, with arguments
lying outside of the permitted band (i.e. the distance of the corresponding
momentum to the Fermi point exceeds $2n_{\rm max}\Delta k_f$). In order to get a
consistent closed set of differential equations, one needs to put these extra
couplings back, inside the set of allowed couplings.

For instance, imagine that $n_{\rm max}=2$, and that $g(3,2,2)$ coupling
intervenes in a RG equation. One cannot, unfortunately, just map
$(3,2,2)\mapsto(2,2,2)$, because these triplets {\em do not belong to the same
symmetry orbit}. In doing so, one would get a very poor $k_\Vert$
dependence; we have actually proved, in the case of a $(p_1,p_2,p'_2)$
representation, that all orbits of $g_0$ except $(0,0,0)$ would collapse into
one single orbit.

Therefore, the truncation procedure {\em must be} compatible with all the
symmetries of the system.  We have used the $(c,l,p)$ notation, which is very
convenient.  One can check that all the symmetries conserve $c+l+p$ modulo
$4\Delta k_f$.  This explains why $n_i\mapsto n_i\pm1$ can't be compatible
with the symmetries.  On the contrary, $n_i\mapsto n_i\pm2$ is completely
compatible, i.e. it maps a triplet onto on an already defined orbit; hence we
have used this mapping for the extra couplings.

With $n_{\rm max}=2$, for each coupling $g$ we find 63 different couplings%
\footnote{that are $\{(i,j,k),i,j,k{=}{-}2,0,2\}$$\bigcup$$\{(\pm1,\pm1,i),
i{=}{-}2,0,2\}$\penalty-10000$\bigcup$$\{(\pm1,i,\pm1),i{=}{-}2,0,2\}$$\bigcup$$\{(i,\pm1,\pm1),
i{=}{-}2,0,2\}$}, which separate into 23 orbits (having 1, 2 or 4 elements),
except for the $g_b$ couplings.  For these, the enumeration is more tedious, we
eventually find 8 orbits (of 4 elements, see~\ref{orbites}).  There are
altogether $3\times23+8=77$ different orbits; if we include the spin separation,
we thus need to calculate 154 coupled differential equations. $n_{\rm max}=3$
gives 390, while $n_{\rm max}=4$ gives 806.

\subsection{Divergences of the susceptibilities}

In the range of values for $U$ that we have investigated, the RG flow is always
diverging.

\begin{figure}[H]
\epsfysize=10cm
\scalebox{0.9}{\rotatebox{-90}{\epsfbox{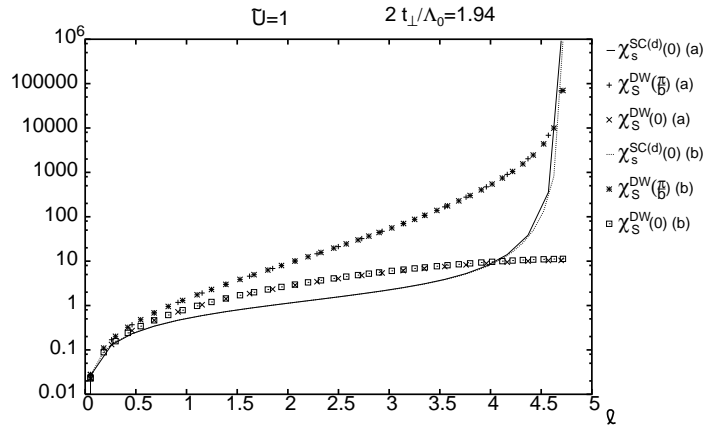}}}
\caption{Flow of the susceptibilities $\chi\lower1.6pt\hbox{${}^{{\rm SC}(d)}_s$}$ and
$\chi^{\rm DW}_S$ (intra or interband), at $\tilde U=1$ and $2t_\perp/\Lambda_0=1.94$~:
(a)~usual RG procedure; (b)~including $k_\Vert$ dependence. You observe
that $\chi^{DW}_{S}(0)$ does not diverge but only reaches a plateau.}
\label{chimixte}
\end{figure}

When the initial interaction Hamiltonian $H_{\rm int}$ is purely local, i.e. when the
interchain scattering are discarded ($C_{\rm back}=C_{\rm for}=0$),
the interband SDW susceptibility $\chi^{\rm DW}_{S}({\pi\over b})$ is always
divergent.  In the superconducting phase, the SC singlet $d$
susceptibility $\chi\lower1.6pt\hbox{${}^{{\rm SC}(d)}_s(0)$}$ is also divergent,
at the same critical scale $\Lambda_c=\Lambda_0\e^{\ell_c}$.  A third
susceptibility $\chi^{\rm DW}_{S}(0)$
increases and reaches a high plateau (see Fig.~\ref{chimixte}).  Almost all
other susceptibilities remain negligible.

When the parameters $C_{\rm back}$ or $C_{\rm for}$ are increased, both
$\chi^{\rm DW}_{S}({\pi\over b})$ and
$\chi\lower1.6pt\hbox{${}^{{\rm SC}(d)}_s$}$ decrease; they are
progressively replaced by the divergence of the CDW susceptibility
$\chi^{\rm DW}_{C}({\pi\over b})$, and, in the case of backward scattering, of
the triplet SC susceptibility $\chi\lower1.6pt\hbox{${}^{{\rm SC}(f)}_t$}(0)$.

\begin{figure*}[t]
\epsfysize=10cm
\rotatebox{-90}{\scalebox{0.22}{\includegraphics{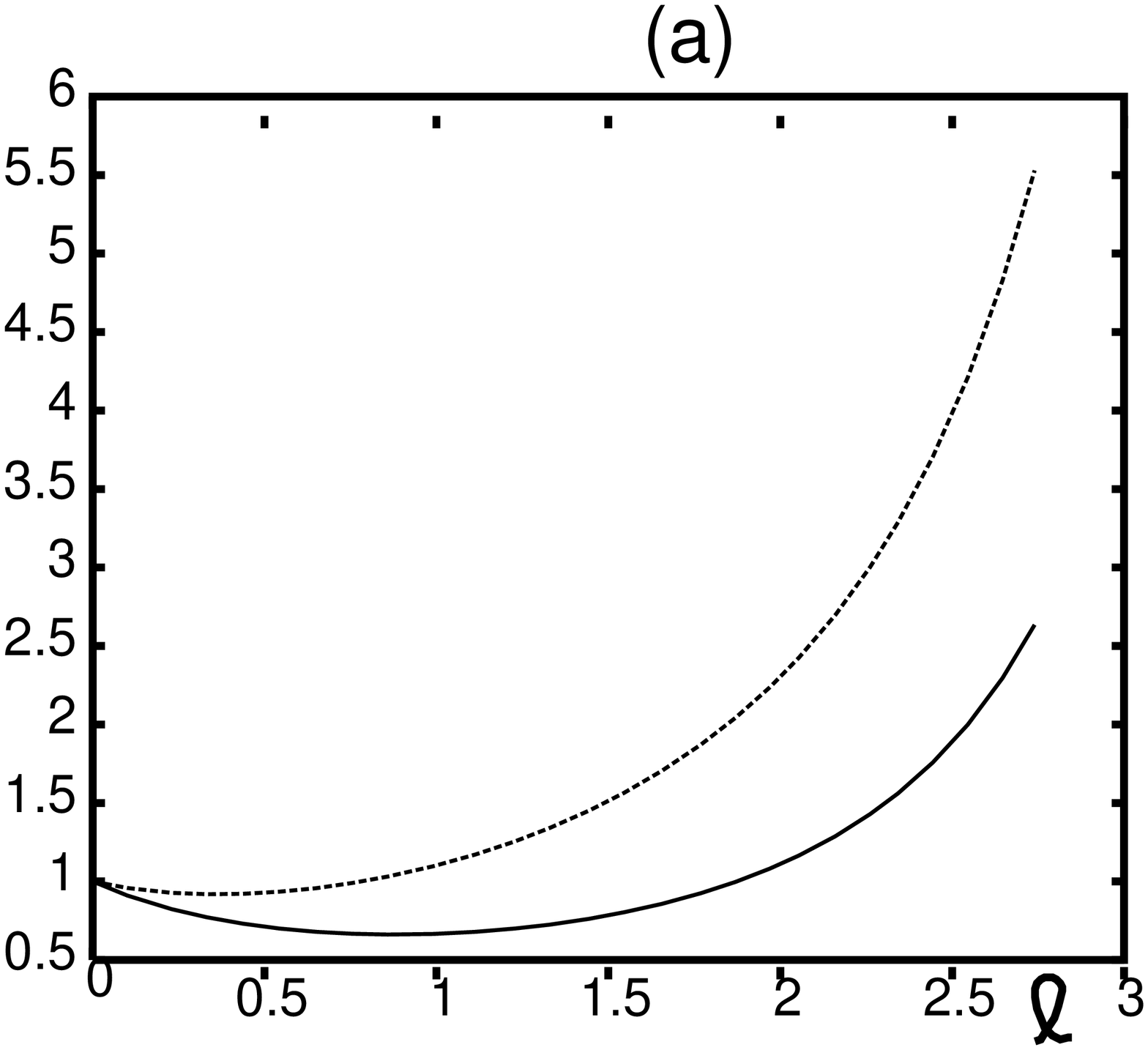}}}
\rotatebox{-90}{\scalebox{0.22}{\includegraphics{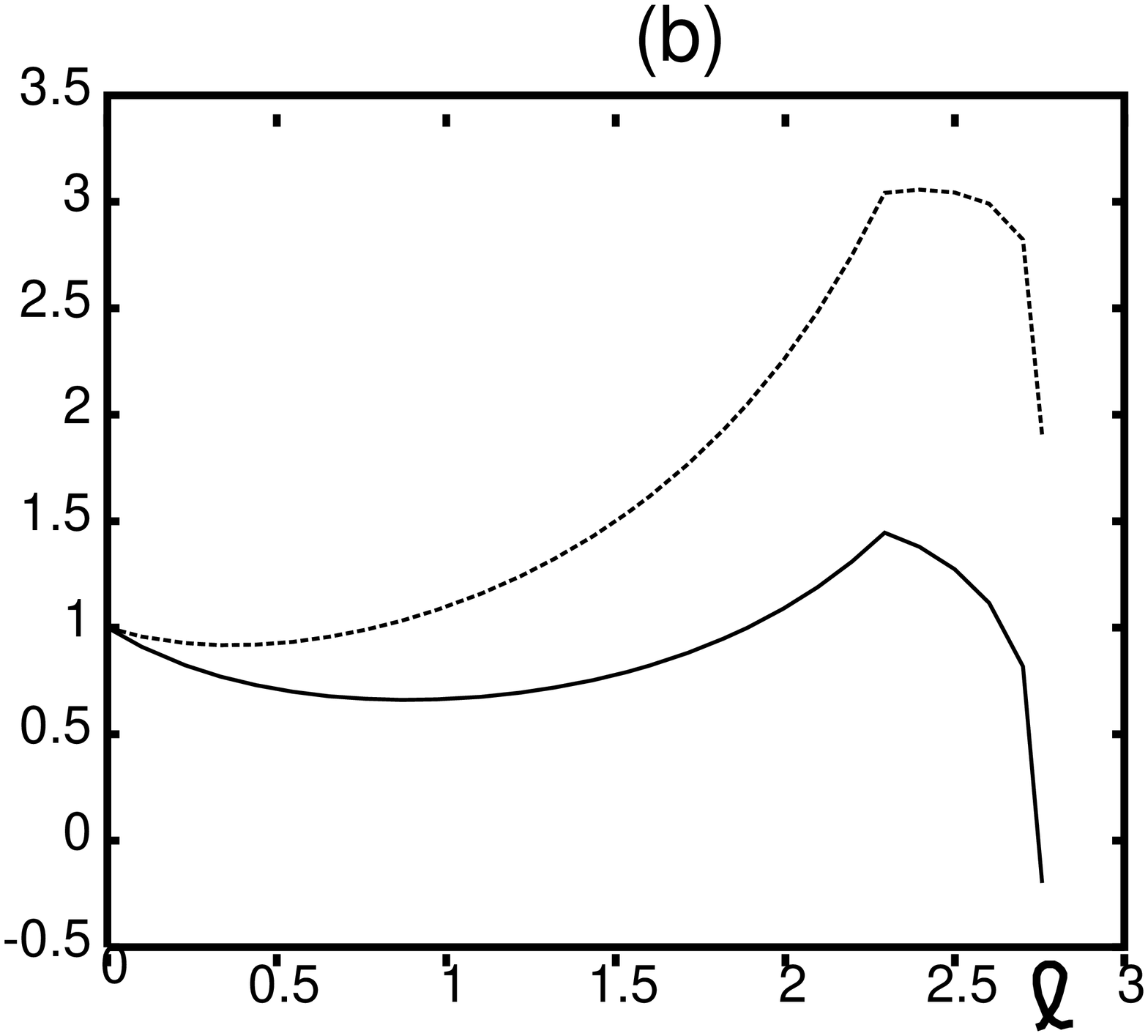}}}
\rotatebox{-90}{\scalebox{0.22}{\includegraphics{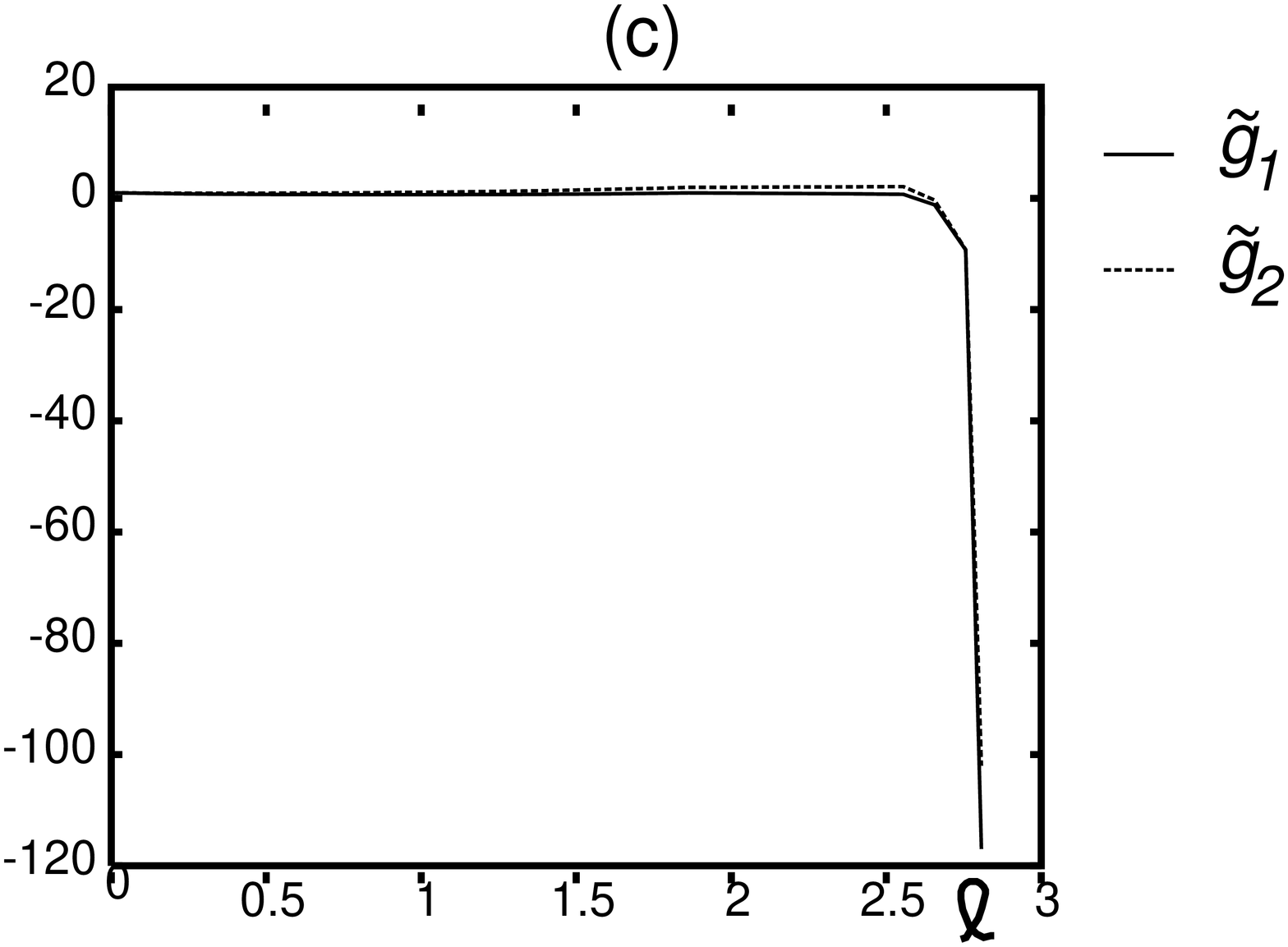}}}
\caption{Flow of the couplings $\tilde g_{01}$ and $\tilde g_{02}$ at~:
(a)~$t_\perp\lesssim t_{\perp c}(U)$;
(b)~$t_\perp=t_{\perp c}(U)$; (c)~$t_\perp\gtrsim t_{\perp c}(U)$}
\label{gchange}
\end{figure*}

One finds at most four divergent susceptibilities
($\chi\lower1.6pt\hbox{${}^{{\rm SC}(d)}_s(0)$}$,
$\chi^{\rm DW}_{C}({\pi\over b})$, 
$\chi\lower1.6pt\hbox{${}^{{\rm SC}(f)}_t$}(0)$ and
$\chi^{\rm DW}_{S}({\pi\over b})$) at a time.

Since the RG flow is diverging, we cannot further calculate the renormalized
couplings.  To deduce a phase diagram, we must find out which mechanism
dominates; we used two different criteria~: according to the first one, we
simply take the susceptibility which reaches the highest value $|\chi|$ at
$\Lambda_c$; according to the second one, we take the susceptibility which has
the highest slope.

These two criteria bring non equal results. Although the first one is a poorer
criterion, its conclusions remain stable when either the precision or
$n_{\rm max}$ are changed. The second is however preferred, as we will see its
conclusion are physically consistent, contrary to the first one.

\section{Phase diagram with initially local interactions}

Let us first discuss the case of initially local interactions
($C_{\rm back}=C_{\rm for}=0$, no interchain scattering). Of course, we can only
fixe $H_{\rm int}$ at $\ell=0$, and the flow will develop non local
interactions.

\subsection{Results}

We begin with the phase diagram obtained when the $k_\Vert$ dependence is
neglected.

\subsubsection{Phase diagram with no $k_\Vert$ dependence}

In the region of the phase diagram that we have investigated ($0\le \tilde U
\le2$), the SC susceptibility $\chi\lower1.6pt\hbox{${}^{{\rm SC}(d)}_s(0)$}$ is
always divergent, as well as the SDW susceptibility
$\chi^{\rm DW}_S({\pi\over b})$ (see Fig.~\ref{chimixte}).

According to the slope criterion,
$\chi\lower1.6pt\hbox{${}^{{\rm SC}(d)}_s(0)$}$ always
dominates. We induce that this region is superconducting (this is consistent
with the conclusions of Fabrizio\cite{Fabrizio}), and that the pairing is of
symmetry $d$; however, the presence of SDW instabilities, developing in
the same region, makes a detailed determination of this phase very uneasy and
beyond the possibilities of our approach. We will call it SC phase.

\subsubsection{Phase diagram with $k_\Vert$ dependence}

\paragraph*{SC phase}

When $t_\perp$ is large, we find similar results. For instance, with
$\tilde U=1$, there is no significant differences for
$2t_\perp/\Lambda_0\ge1.94$ (see Fig.~\ref{chimixte}).

\paragraph*{SDW phase}

\begin{figure}[H]
\epsfysize=10cm
\rotatebox{-90}{\epsfbox{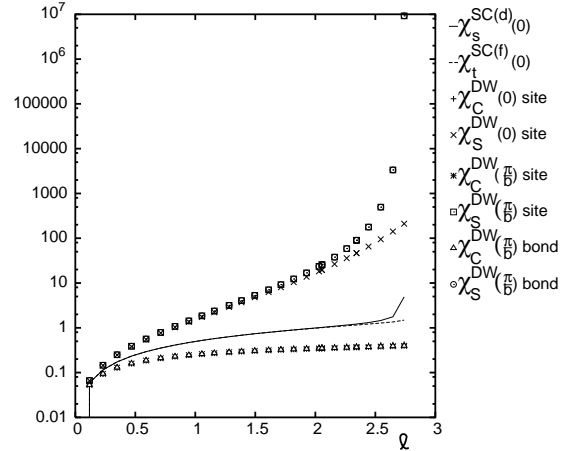}}
\caption{Flow of some susceptibilities, at $\tilde U=1$ and $2t_\perp/\Lambda_0=0.016$.
$\chi\lower1.6pt\hbox{${}^{{\rm SC}(d)}_s(0)$}$ is almost suppressed at $\Lambda_c$.}
\label{chiSDW}
\end{figure}

On the contrary, the superconducting susceptibility is almost suppressed  when
$t_\perp$ is small enough. For $\tilde U=1$ and $2t_\perp/\Lambda_0=0.016$,
$\chi\lower1.6pt\hbox{${}^{{\rm SC}(d)}_s(0)$}$ is  5 orders of magnitude smaller than
$\chi^{\rm DW}_S({\pi\over b})$, at $\Lambda_c$, see on Fig.~\ref{chiSDW}.
In this phase, the SDW 's instability develops so rapidly that it overwhelms
all other processes. This indicates indeed a pure SDW phase.

Hence, this result differs drastically from those obtained when the $k_\Vert$
dependence is neglected.

Moreover, we observe a transition between the SDW phase and the SC one, at a
critical parameter $t_{\perp c}(U)$.

\paragraph*{Critical behavior}

We characterize this transition by different ways.

First of all, the behaviour of the renormalized two-particle couplings change
very rapidly, at $t_{\perp c}(U)$: as $t_\perp$ decreases,
$|\tilde g_{01}(\Lambda_c)|$ and $|\tilde g_{02}(\Lambda_c)|$ shrink suddenly,
then, after a little interval, $\tilde g_{01}(\Lambda_c)$ and
$\tilde g_{02}(\Lambda_c)$ become positive (and even $>\tilde U$, see on
Fig.~\ref{gchange}). We also observe changes, though less significant, for the
other couplings $\tilde g$ (for instance, $|\tilde g_{f01}(\Lambda_c)|$
decreases and $\tilde g_{b1}(\Lambda_c)$ becomes negative when $t_\perp$
increases).

\begin{figure}[H]
\epsfysize=10cm
\rotatebox{-90}{\epsfbox{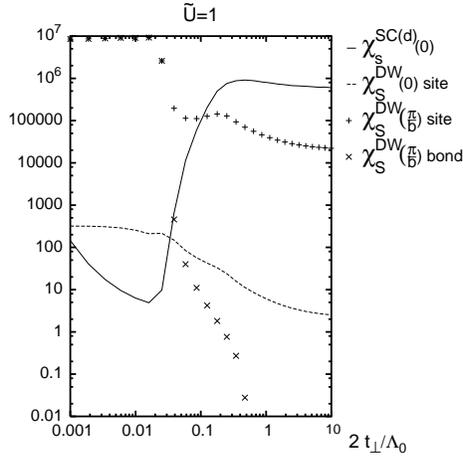}}
\caption{Curves of $\chi\lower1.6pt\hbox{${}^{{\rm SC}(d)}_s$}(\Lambda_c)$ and
$\chi^{\rm DW}_S(\Lambda_c)$ versus $2t_\perp/\Lambda_0$.}
\label{chifinal}
\end{figure}

Moreover, we observe (on Fig.~\ref{chifinal}) a marked site/bond separation of
the SDW susceptibility, at $t_{\perp c}(U)$. The site and bond SDW
susceptibilities are degenerate for $t_\perp\le t_{\perp c}(U)$ (in the SDW
phase), which is consistent with the fact they should be equal at $t_\perp=0$
(where the system is purely one-dimensional, see Ref.~\cite{Bourbonnais}), while they
smoothly separate after the transition. The same site/bond separation occurs for
the CDW susceptibility.

\paragraph*{Transition region}

The behaviour of most of the parameters that we have examined indicates the same
critical value $t_{\perp c}(U)$, which we have determined exactly, using the
slope criterion.

However,$|\chi\lower1.6pt\hbox{${}^{{\rm SC}(d)}_s$}(0)|
<|\chi^{\rm DW}_S({\pi\over b})|$ holds until $t_\perp$ reaches a value
$t_{\perp c2}(U)$; this second critical value, which corresponds to the height
criterion, is confirmed by minor modifications of behaviour, which occur in the
interval $t_{\perp c}(U)<t_\perp\le t_{\perp c2}(U)$ and are very smooth (for
instance, $\tilde g_{b1}(\Lambda_c)$ and $\tilde g_{b2}(\Lambda_c)$ cross) .

\begin{figure}[H]
\epsfysize=10cm
\rotatebox{-90}{\epsfbox{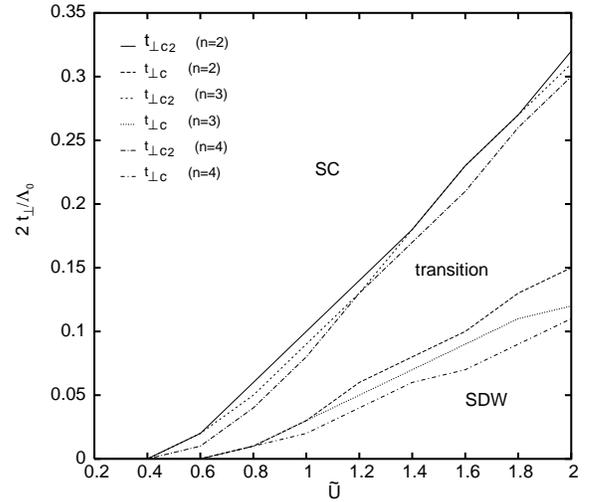}}
\caption{Phase diagram when the $k_\Vert$ dependence is included.}
\label{phasetot}
\end{figure}

The numerical determination of $t_{\perp c2}(U)$ is very stable (see
Fig.\ref{phasetot}), and the complete behaviour, from the SDW phase to the SC
phase, is clear on Fig.~\ref{chifinal}, which shows the absolute values
of the susceptibilities at $\Lambda_c$.

The region $t_{\perp c}(U)<t_\perp\le t_{\perp c2}(U)$ is called transition
region, we believe it is a superconducting phase, where SDW instabilities
seem however to dominate.  The SDW are precursor manifestations
of the pure SDW phase, which is next to the transition region.

As already stated, our results converge very rapidly when the band width on
which we project the momenta is increased.  The complete phase diagram (for
$C_{\rm back}=C_{\rm for}=0$) is shown on Fig.~\ref{phasetot}, for
$n_{\rm max}=2,$ 3 and 4.

\begin{figure}[H]
\epsfysize=10cm
\rotatebox{-90}{\epsfbox{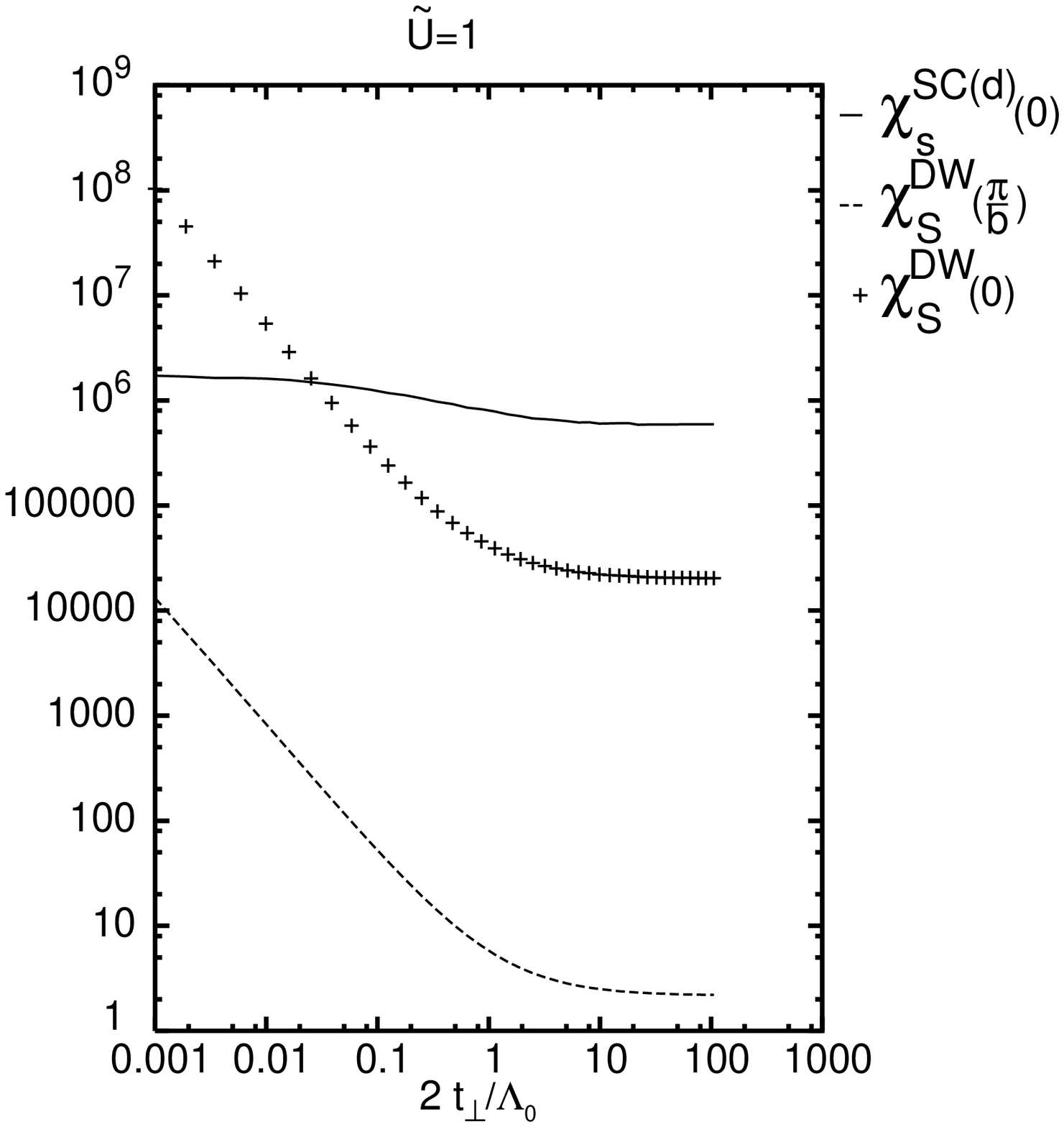}}
\caption{Curves of $\chi\lower1.6pt\hbox{${}^{(d)}_s$}(\Lambda_c)$ and
$\chi^{\rm DW}_S(\Lambda_c)$ versus $2t_\perp/\Lambda_0$ when
$k_\Vert$ is neglected.}
\label{transitionproj}
\end{figure}

Moreover, it is most interesting to note that, contrary to the pure SDW phase,
the transition region can be detected when the $k_\Vert$ dependence is
neglected (the value of $t_{\perp c2}(U)$ is lowered).  This can be seen, for
instance, on Fig.~\ref{transitionproj}, which corresponds to Fig.~\ref{chifinal}
with no $k_\Vert$ dependence.

\paragraph*{SC critical temperature}

\begin{figure}[H]
\epsfysize=10cm
\scalebox{0.8}{\rotatebox{-90}{\epsfbox{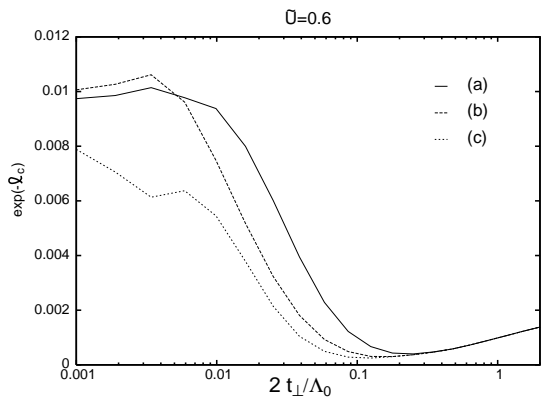}}}
\caption{Curves of $\Lambda_c/\Lambda_0=\e^{-\ell_c}$ versus
$2t_\perp/\Lambda_0$ with $n_{\rm max}=2$~(a), $n_{\rm max}=3$~(b) and
$n_{\rm max}=4$~(c).}
\label{tcsupra}
\end{figure}

The critical energy $\Lambda_c$ at which SC susceptibility diverges gives
an approximate indication of the SC critical temperature. We give
a plot of $\Lambda_c/\Lambda_0$ versus $2t_\perp/\Lambda_0$; as seen on
Fig.~\ref{tcsupra}, $\Lambda_c$ is roughly decreasing with $t_\perp$. The
band gap parameter $t_\perp$ is also increasing with pressure; therefore,
this behaviour is compatible with the experimental data, which show that $T_c$
is decreasing with pressure in quasi-one-dimensional organic compounds.

\subsection{Discussion}

As already mentioned, as soon as the dependence of $k_\Vert$ is included, we
observe two separated phases, one purely SDW, the other one a SC phase with
competing SDW instabilities.

On the other hand, for $t_\perp\ge1$, i.e. when the initial bandwidth lies
inside $[k_{f\pi},k_{f0}]$, the $k_\Vert$ dependence has no observable
influence on the susceptibilities.

In the SDW phase, our results prove the existence of large antiferromagnetic
fluctuations. We believe that these SDW instabilities are not the signature of
a localized antiferromagnetic ground state, but of antiferromagnetic itinerant
electrons, as it is indeed observed in Bechgaard salts. Actually, the flow is
driving towards a fixed point, which does not seem to be the one-dimensional
solution~: for instance, the renormalized couplings $\tilde g_1$ and $\tilde
g_2$ of the 1-$d$ solution are 0 and 1/2 and differ from the values which we
obtain when the flow is diverging, in the SDW phase (see Fig.~\ref{gchange} (a)).

We induce that the spin-gap should disappear in this SDW region, which is
consistent with what Park and Kishine\cite{Park,Kishine2} claim.

In the SC phase, the SC divergence is due to the Cooper channel, while
that of density waves is due to the Peierls one (see, in the case of a single
band model, Refs.~\cite{Prigodin,Solyom}). The appearance of $d$-wave
superconductivity in ladder systems is well understood within a strong
coupling scenario, where a spin gap leads to interchain Cooper pairing. However,
in our calculations, we see that superconducting correlations are always
enhanced by SDW fluctuations.
Contrary to what Lee {\it et al.} claimed first\cite{Lee}, there is an itinerant
electron mechanism in this case, which is the weak coupling equivalent of the
localized electron mechanism in  the strong electron scenario. It was proposed
by Emery\cite{Emery}, and is essentially the spin analog of Kohn-Luttinger
superconductivity. The mutual enhancement of the two channels is also discussed
in Refs.~\cite{Honerkamp2,Furukawa}.

As a consequence of this mutual enhancement, the spin-gap should not appear
with the first appearance of SC instabilities, but for somehow larger values of
$t_\perp$.

Moreover, we observe that the SC pairing is a $\bf Q=0$ mechanism, while the SDW
are excited by ${\bf Q}=(2\Delta k_f,{\pi\over b})$ vectors.  This can be
explained by the symmetry of each channel.  The Green function of the Cooper
channel gives a factor $1/(\epsilon({\bf k})+\epsilon(-{\bf k}))$ and is
minimized with the ${\bf k}\mapsto-{\bf k}$ symmetry, which corresponds to an
intraband process.  The Green function of the Peierls channel gives a factor
$1/(\epsilon({\bf k})+\epsilon({\bf k+Q}))$ and is minimized with the
${\bf k}\mapsto{\bf k+Q}$ symmetry, which corresponds to an interband process.

This can also be seen in the RG equations. $d\ln(z^{\rm SC}(0))/d\ell$ depends
only on $g_0$ and $g_t$, whereas $d\ln(z^{\rm SC}({\pi\over b}))/d\ell$ depends
on $g_f$ and $g_b$. Since $g_b$ processes  are depressed as soon as $\Lambda\le2
t_\perp$, 0-condensate are favored. Moreover, this predominance is stabilized
by the $dg_t/d\ell$ equations, in which the Cooper term depends on $g_0$, and by
the $dg_b/d\ell$ equations, in which the Cooper term depends on $g_f$.

The same argument applies for DW instabilities. $d\ln(z^{\rm DW}(0))/d\ell$
depends on $g_0$ and $g_b$, whereas $d\ln(z^{\rm DW}({\pi\over b}))/d\ell$
depends on $g_f$ and $g_t$. So, $\pi$-processes are favored. Again, this
is stabilized by the $dg_t/d\ell$ equations, in which the Peierls term depends
on $g_f$, and by the $dg_b/d\ell$ equations, in which the Peierls term depends
on $g_0$.

The critical temperature, in the SC phases, is indicated in Fig.~\ref{tcsupra}.
We chose $\tilde U=0.6$ in order to avoid the SDW phase. The general trend is
that of a quasi-one-dimensional system; the increasing curve, for small
values of $t_\perp$, can be related to transition effect and to the furthered
influence of the SDW fluctuations.

\section{Phase diagram with initial Coulombian interchain scattering}

\subsection{Results}

\subsubsection{Influence of a backward interchain scattering}

Let us now study the effect of a backward interchain scattering. This type of
coupling has been inverstigated by Bourbonnais {\it et al.}\cite{Bourbonnais2} in the
context of correlated quasi-onde-dimensional metals, for which CDW correlations
are enhanced and triplet superconducting instabilities can occur. When the
parameter $\tilde C_{\rm back}$ is increased, the behaviour of the
susceptibilities depends on the parameters $(t_\perp,\tilde U)$.

\paragraph*{Appearance of triplet SC and CDW}

\begin{figure*}[t]
\epsfysize=10cm
\rotatebox{-90}{\scalebox{0.39}{\includegraphics{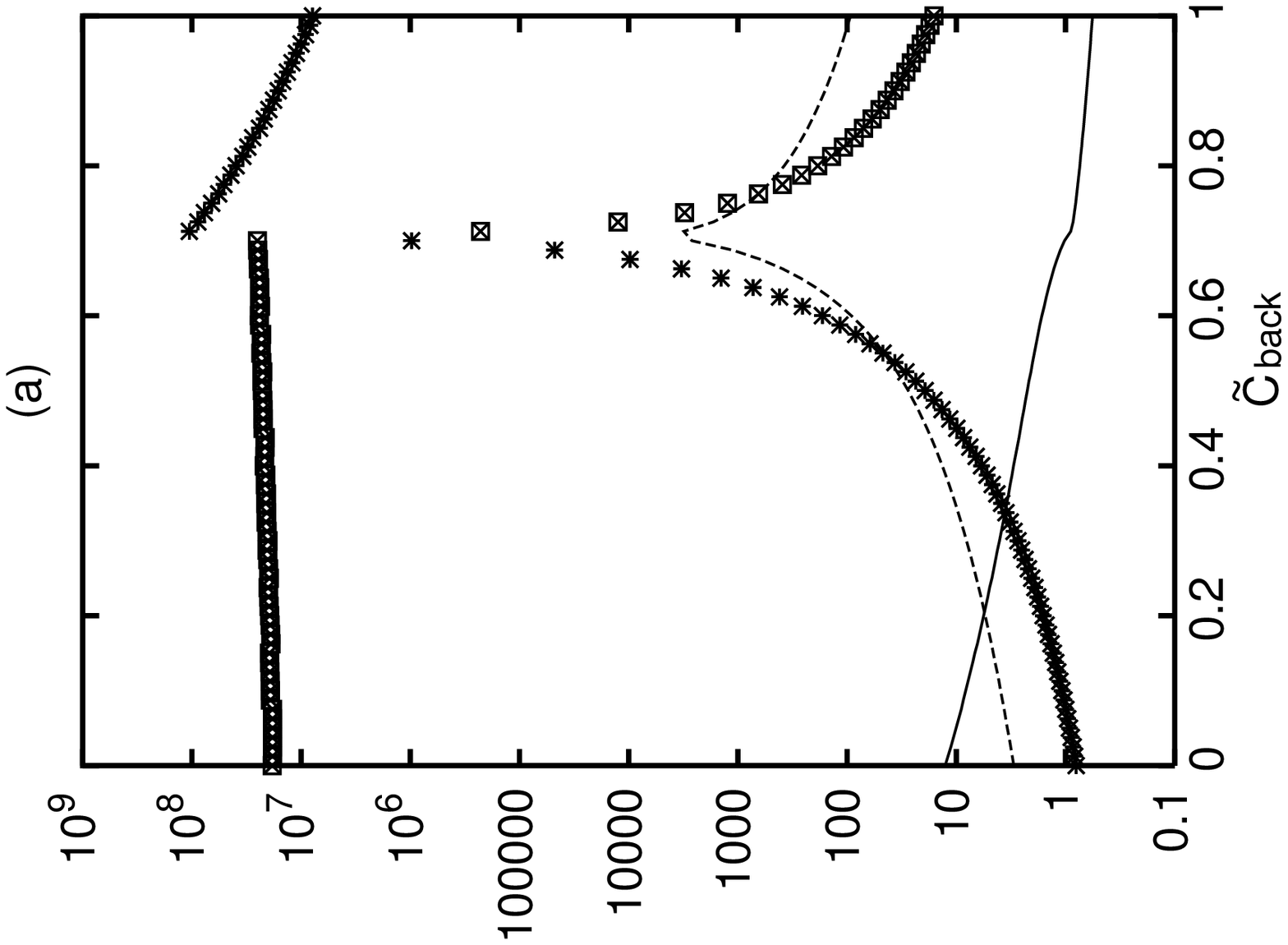}}}
\hspace{-4cm}
\rotatebox{-90}{\scalebox{0.39}{\includegraphics{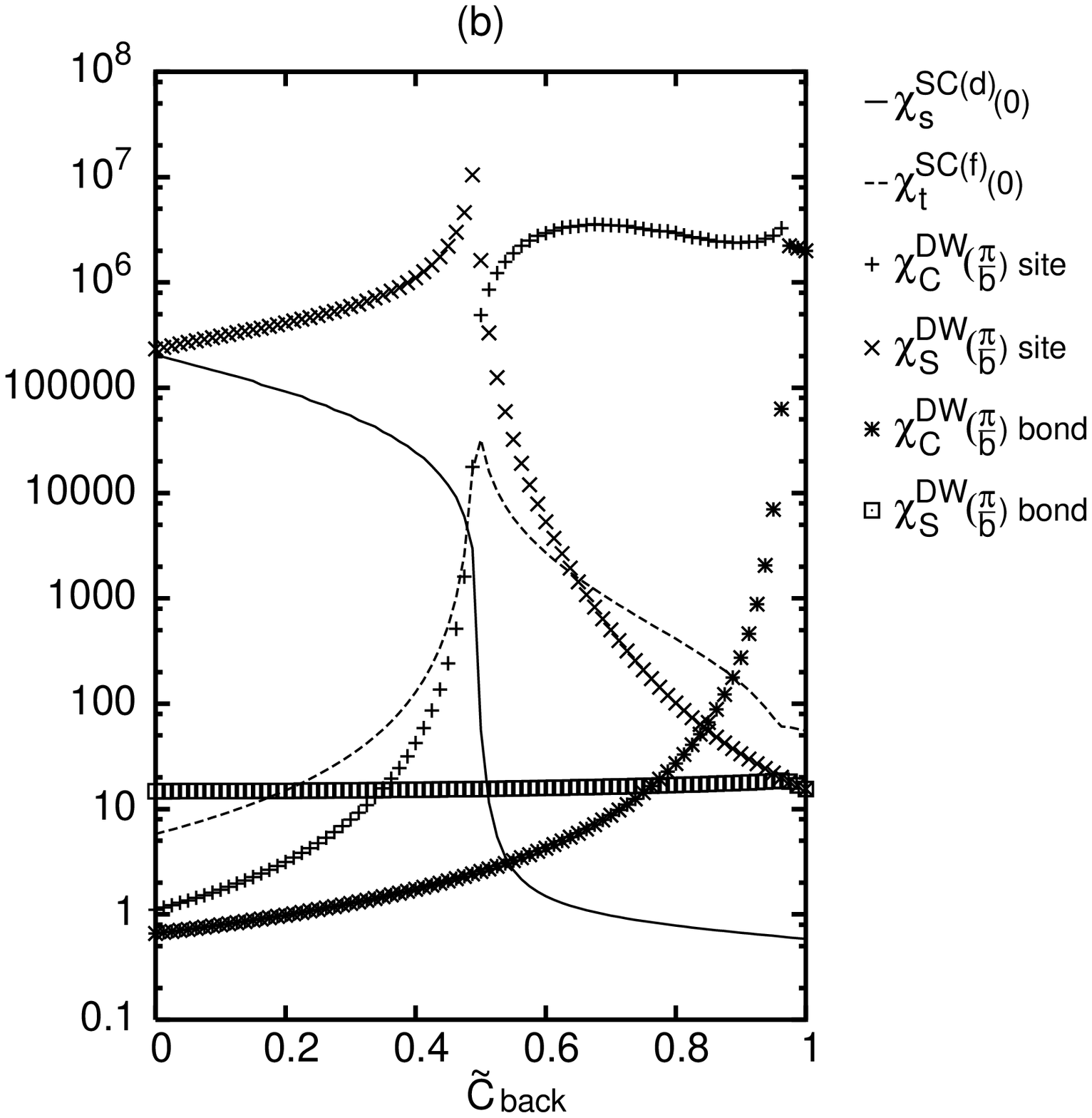}}}
\caption{Curves of the $\chi(\Lambda_c)$ versus $\tilde C_{\rm back}$, at
$\tilde U=1$ and a)~$2t_\perp/\Lambda_0=0.01$; b)~$2t_\perp/\Lambda_0=0.1$.}
\label{coulombS}
\end{figure*}

\begin{figure}[H]
\epsfysize=10cm
\scalebox{0.8}{\epsfbox{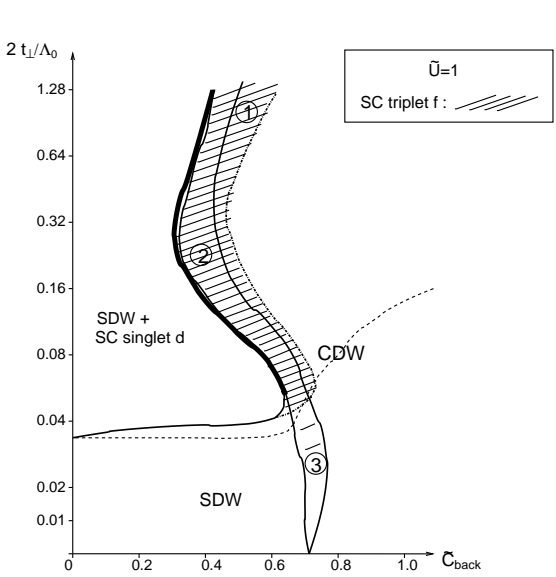}}
\caption{Phase diagram for $\tilde U=1$. The shaded area indicates the
divergence of the triplet susceptibility. The dashed line separates site/bond
degenerated states (below) and non degenerated ones (above). Other lines and
domains are explained in the legend or in the text.}
\label{phasediag}
\end{figure}

For $t_\perp\le t_{\perp c}(U)$, the SDW phase exists for $\tilde C_{\rm back}$
small enough.  As $\tilde C_{\rm back}$ is further increased, the SDW
instabilities are replaced by CDW ones. The transition is smooth, and there is
a narrow region where both SDW and CDW coexist (region $\circled3\;$ in
Fig.~\ref{phasediag}). We show a $2t_\perp/\Lambda_0=0.01$ section of the susceptibilities
at $\Lambda_c$ on Fig.~\ref{coulombS}.

\begin{figure}[H]
\epsfysize=10cm
\rotatebox{-90}{\epsfbox{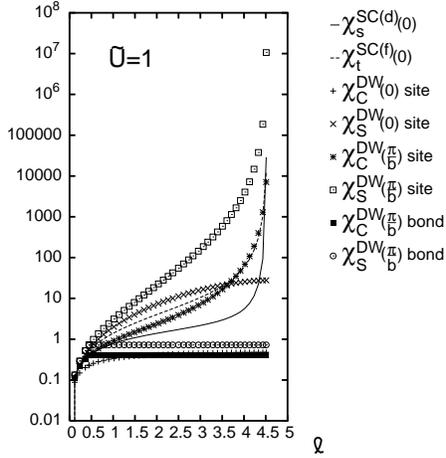}}
\caption{Flow of the susceptibilities for $2t_\perp/\Lambda_0=0.32$ and
$\tilde C_{\rm back}=0.18$.}
\label{4chi}
\end{figure}

For $t_\perp\ge t_{\perp c}(U)$, the SC phase (with SC singlet $d$ and SDW
instabilities) exists for $\tilde C_{\rm back}$ small enough. As $\tilde C_{\rm
back}$ is further increased, the singlet SC modes are replaced by triplet ones,
while SDW are replaced by CDW. Singlet and triplet SC appear to be antagonistic,
and the transition is very pronounced; in the coexistence line between them, one
also finds SDW and CDW divergences (see Fig.~\ref{4chi}). On the contrary, the
transition between SDW and CDW is very smooth, although the coexistence region
is still narrow (region $\circled2\;$ in Fig.~\ref{phasediag}). We show a
$2t_\perp/\Lambda_0=0.1$ section of the susceptibilities at $\Lambda_c$ on
Fig.~\ref{coulombS}.

The triplet SC condensate has $f_x$ symmetry. The corresponding susceptibility
is mostly divergent in a region of coexistence with SDW and CDW (region
$\circled2\;$ in Fig.~\ref{phasediag}), but it is also divergent in a region of
coexistence with only CDW (region $\circled1\;$ in Fig.~\ref{phasediag}).

\begin{figure*}[t]
\epsfysize=10cm
\rotatebox{-90}{\scalebox{0.36}{\includegraphics{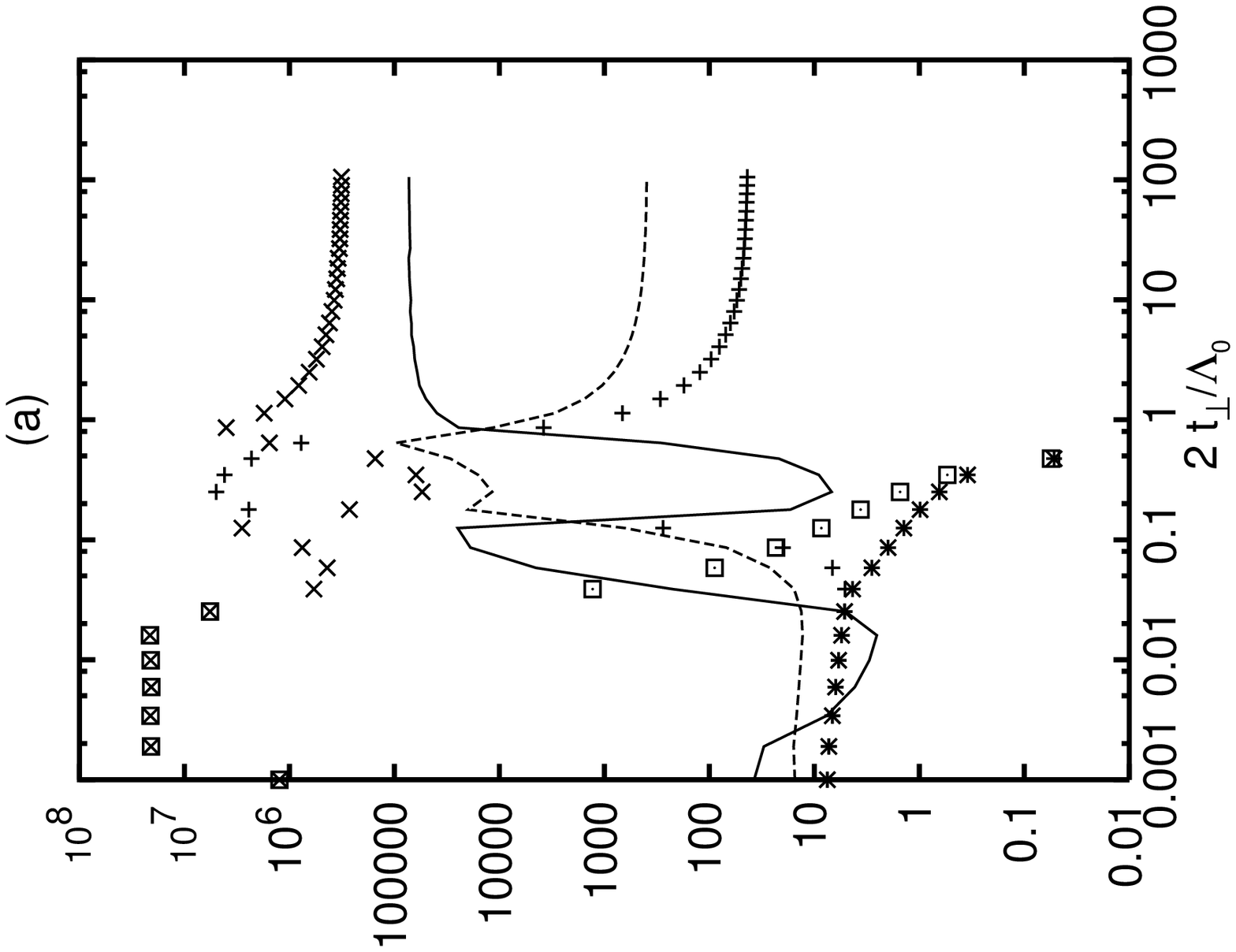}}}
\hspace{-1cm}
\rotatebox{-90}{\scalebox{0.37}{\includegraphics{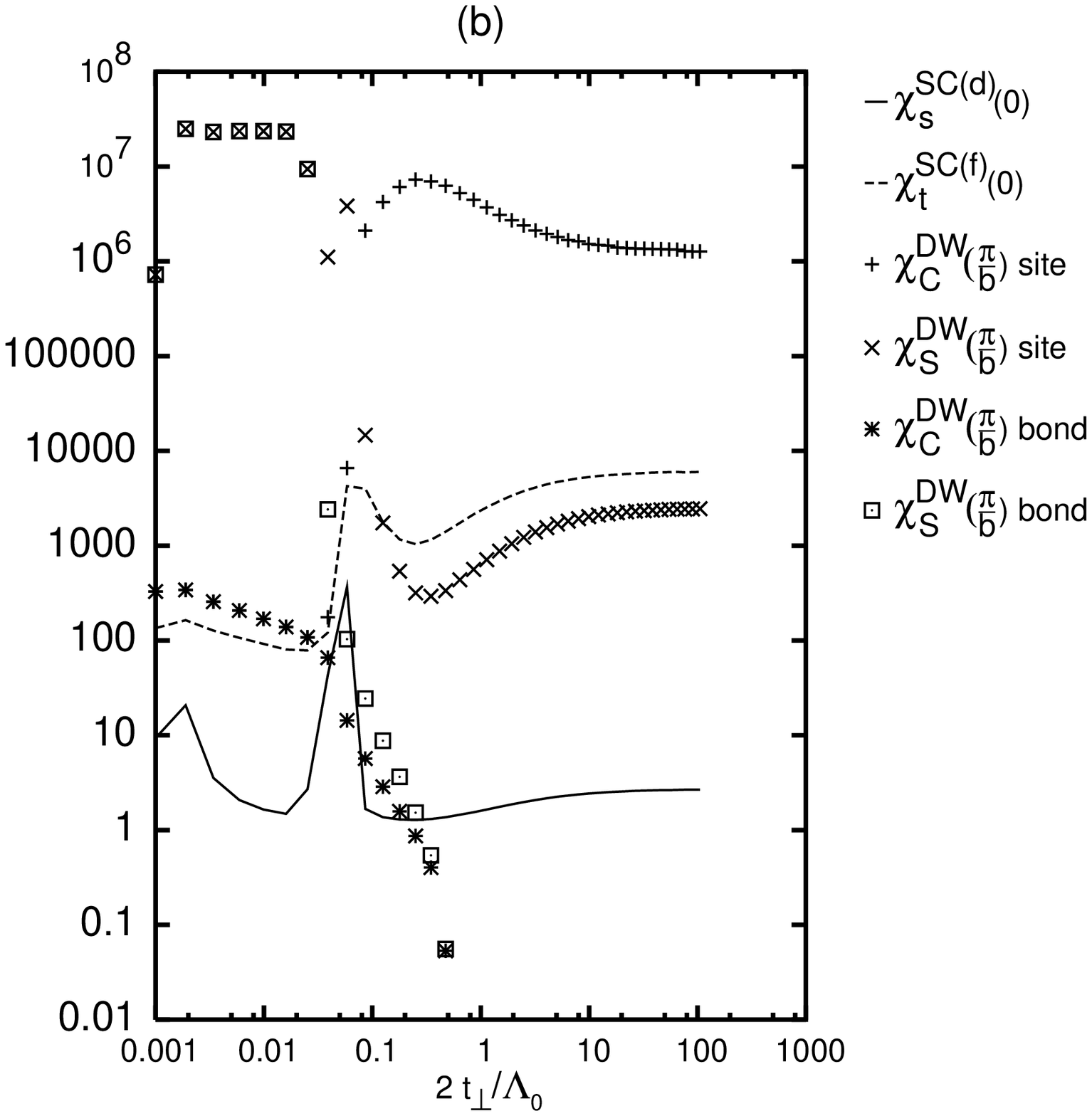}}}
\caption{Curves of the susceptibilities $\chi(\Lambda_c)$ versus
$2t_\perp/\Lambda_0$, for~: a)~$\tilde C_{\rm back}=0.4$;
b)~$\tilde C_{\rm back}=0.6$.}
\label{coulomb}
\end{figure*}

When $\tilde C_{\rm back}$ is large enough, the triplet SC modes are suppressed,
and the region is a pure CDW phase. We show a section of the susceptibilities at
$\Lambda_c$, for $\tilde C_{\rm back}=0.2$ and $\tilde C_{\rm back}=0.3$, on
Fig.~\ref{coulomb}, which clearly indicates the domain of existence of the
triplet SC.

\paragraph*{Site/bond separation}

As we have already observed it, in the case of $C_{\rm back}=C_{\rm for}=0$, for
small values of $t_\perp$, site and bond SDW susceptibilities are degenerate,
as well as site and bond CDW ones.

This generalizes for all values of $\tilde C_{\rm back}$. The site/bond
separation line is an increasing function $t_{\perp g}(C_{\rm back})$ of
$C_{\rm back}$, shown on Fig.~\ref{phasediag}; For small values of $U$, this
line crosses the SC domain, but for $\tilde U=1$ it is already disconnected
from the SC frontier (although it remains close to it).

\subsubsection{Influence of a forward interchain scattering}

The phase diagram when $C_{\rm for}$ is included is very rich, and beyond the
scope of this article.

\begin{figure}[t]
\epsfysize=10cm
\rotatebox{-90}{\epsfbox{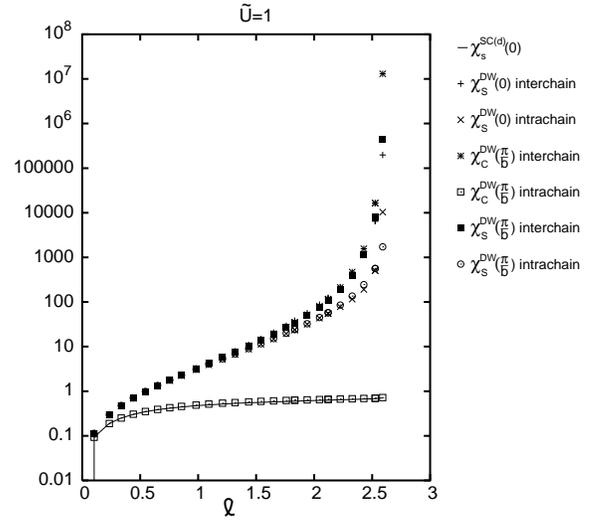}}
\caption{Flow of the susceptibilities, $t_\perp=0.02$ and~:
$\tilde C_{\rm for}=0.95$.}
\label{cavant}
\end{figure}

We would like to emphasize only the fact that all SC instabilities are
suppressed when $C_{\rm for}$ is increased. Fig.~\ref{cavant} gives a typical
flow of the susceptibilities, with a large $\tilde C_{\rm for}$.

\subsection{Discussion}

Let us analyze these behaviors, which follow simple trends.

The CDW instabilities are enhanced when $g_C$ is increased, whereas SDW ones are
enhanced when $g_S$ is increased (this can be verified in the corresponding RG
equations of \ref{eqRGz}). Similarly, singlet SC instabilities are enhanced
when $g_s$ is increased, whereas triplet ones are enhanced when $g_t$ is
increased.

So, an increase of $C_{\rm back}$ implies an increase of the real space
coupling $g_1$, and thus, from Eq.~(\ref{relgSDW}), it favors CDW instabilities
against SDW ones, and from Eq.~(\ref{relgSC}), it favors triplet SC
instabilities and depresses singlet SC ones.

Similarly, an increase of $C_{\rm for}$ implies an increase of the real space
coupling $g_2$, and thus, from Eq.~(\ref{relgSDW}), it favors CDW and SDW
instabilities, and from Eq.~(\ref{relgSC}), it depresses SC ones.

Of course, we examine here the influence of parameters $C_{\rm back}$ and
$C_{\rm back}$ on the bare couplings. However, we believe that the flow could
not just simply reverse this influence, even if the renormalized values of the
couplings differ a lot from their bare values. Moreover, one verifies that these
conclusions exactly correspond to the observed behaviour.

The density wave interactions are on site, whereas the SC pairing are
inter-site (except for singlet $s$ one), so we believe that the DW instabilities
appear first, and then enhance the SC ones. This is not true of the DW bond
correlations, but we observed no divergences of these ones, and we have
not studied any other sophisticated inter-site DW excitation response.

From this point of view, the fact that $\pi$ DW processes are favored, as we
already discussed before, implies a $\pi$ dephasing between both chains of the
ladder. The $\pi$ dephasing of the SDW thus fits perfectly singlet $d$
condensate (which consists in a pairing of two electrons on a rung, with
opposite spins, see Fig.~\ref{spairing} (b)). This accounts nicely for the
appearance of singlet SC instability, induced by SDW one.

In the same trend of ideas, the $\pi$ dephasing of the CDW fits triplet $f_x$
condensate (which consists in a pairing of two electrons on each chain, one
stepped by unity from the other, see Fig.~\ref{tpairing} (a)) and accounts for the
appearance of triplet SC instability, induced by CDW one.

On the contrary, the triplet $p_x$ condensate consists in two following electrons
on one chain (see Fig.~\ref{tpairing} (b)), this pairing is not enhanced by CDW
instabilities; in fact, it is the analog of singlet $s$ condensate, which is not
either enhanced by SDW instabilities, and is therefore disadvantaged, compared
to $d$ pairing.

As can be observed on Fig.~\ref{tpairing}, triplet $f_x$ condensate are not
incompatible with CDW. For instance, one could easily figure out a succession
of condensate, with alternate spins, inducing back a global modulation of the
chains. A similar scenario is not possible with triplet $p$ condensate.

One should be aware that the symmetry classification we have used is very
specific of the ladder system, and could not be extended to an infinite number
of chains. The difference between $p$ and $f$ condensate is very subtle and the
situation could reveal quite different in the general quasi-one-dimensional
systems.

\section{Conclusion}

We have investigated the phase diagram of a ladder system, in the Hubbard model,
with an interchain coupling $t_\perp$, using functional RG
method, in the OPI scheme.  We have introduced an original parameterization of
the $k_\Vert$ dependence, and obtained rather new results, in particular, we
have proved the existence of a new phase with only SDW fluctuations,
for small enough values of $t_\perp$. From the divergences of the scattering
couplings, we induce that this phase is different from the one-dimensional
solution. However, for very small values of $t_\perp$ ($t_\perp<\Lambda_0
10^{-4}$), we find the usual one-dimensional behaviour.

Our results altogether prove that the $k_\Vert$ dependence is important
and must be taken into account in such a ladder system. The fact that this
variables become influential in a ladder does not mean that the corresponding
couplings $g(2\Delta k_f,-2\Delta k_f,0)$, etc., are relevant. In fact, if the
cut-off $\Lambda\to0$, these couplings are left out of the integrated band, so
they could only be marginal\cite{Shankar}. However, the divergence takes place
at $\Lambda_c$, which is of the order of $\Delta k_f$, and this explains why
these couplings, which are shifted by $\pm2\Delta k_f$ from the Fermi points,
have a non trivial behaviour and have to be taken into account. Moreover, as
already explained in~\ref{dependance}, during the flow, these couplings
influence those, with all momenta at the Fermi points, until $\Lambda=2\Delta
k_f$. This influence is still sensitive, when the divergence takes place. This
explains why we could distinguish a new phase, which has not yet been observed
by usual methods.

When $t_\perp$ is very large, however (for instance, $t_\perp\sim\Lambda_0$),
the flow continues up to $\Lambda_c\ll v_f\Delta k_f$ (otherwise, the
integrated band would not vary much and the renormalized couplings neither
differ much from their bare values), and the above argument applies,
proving that couplings $g(2\Delta k_f,-2\Delta k_f,0)$, etc., are marginal or
irrelevant. In that case, $k_\Vert$ are not influential, and our results
coincide indeed with former calculations.

We have also given a detailed classification of the response function, which
provides a convenient tool for the determination of order parameters and of
related susceptibilities, corresponding to different instability processes.

We are proceeding now to a complete study of the long range correlations, and
in particular, of the uniform susceptibility. This task however proves  quite
difficult, because of the $k_\Vert$ dependence, which has to be carefully
taken into account. We expect that the spin-gap will indeed disappear in the
SDW phase we have brought to evidence.

We have also investigated the influence of interchain scattering, and showed
that a backward interchain scattering can raise triplet superconductivity, a
resulst consistent with the conclusions of a previous work by Bourbonnais
{\it et al.}\cite{Bourbonnais2,Nickel2} on correlated quasi-one-dimensional
metals. The appearance of triplet SC in a ladder is a very exciting and
promising result, since various authors\cite{Naughton,Cherng} claim to have
found experimental evidence of these instabilities. Even the narrowness of
the triplet SC existence region seems to fit the experimental data, which report
high sensitivity of these fluctuations to some key parameters. This work gains
to be compared with the previous work of Varma {\it et al.}, who did similar
investigations\cite{Varma}.

\vskip0.5cm
We would like to thank N. Dupuis, S. Haddad and B. Douçot for fruitful
discussions and advices. J.~C.~Nickel wishes to thank the Gottlieb Daimler- und
Karl Benz-Stiftung for partial support.

\section*{Appendix}
\appendix
\renewcommand{\thesubsection}{\Alph{subsection}}
\subsection{Couplings}

\subsubsection{Two-particle couplings}
\label{defcouplings}

Here are the definitions of the different couplings $g$, from the two-particle
parameter $\cal G$, and the corresponding diagrams.
$$
\begin{array}{c}
g_0(p_1,p_2,p'_2,p'_1)=\\
{\cal G}(k_{f0}+p_1,-k_{f0}+p_2,-k_{f0}+p'_2,k_{f0}+p'_1)\\
g_\pi(p_1,p_2,p'_2,p'_1)=\\
{\cal G}(k_{f\pi}+p_1,-k_{f\pi}+p_2,-k_{f\pi}+p'_2,k_{f\pi}+p'_1)\\
g_{\!f0}(p_1,p_2,p'_2,p'_1)=\\
{\cal G}(k_{f0}+p_1,-k_{f\pi}+p_2,-k_{f\pi}+p'_2,k_{f0}+p'_1)\\
g_{\!f\pi}(p_1,p_2,p'_2,p'_1)=\\
{\cal G}(k_{f\pi}+p_1,-k_{f0}+p_2,-k_{f0}+p'_2,k_{f\pi}+p'_1)\\
g_{t0}(p_1,p_2,p'_2,p'_1)=\\
{\cal G}(k_{f0}+p_1,-k_{f0}+p_2,-k_{f\pi}+p'_2,k_{f\pi}+p'_1)\\
g_{t\pi}(p_1,p_2,p'_2,p'_1)=\\
{\cal G}(k_{f\pi}+p_1,-k_{f\pi}+p_2,-k_{f0}+p'_2,k_{f0}+p'_1)\\
g_{b0}(p_1,p_2,p'_2,p'_1)=\\
{\cal G}(k_{f0}+p_1,-k_{f\pi}+p_2,-k_{f0}+p'_2,k_{f\pi}+p'_1)\\
g_{b\pi}(p_1,p_2,p'_2,p'_1)=\\
{\cal G}(k_{f\pi}+p_1,-k_{f0}+p_2,-k_{f\pi}+p'_2,k_{f0}+p'_1)
\end{array}
 $$
\vglue-1cm
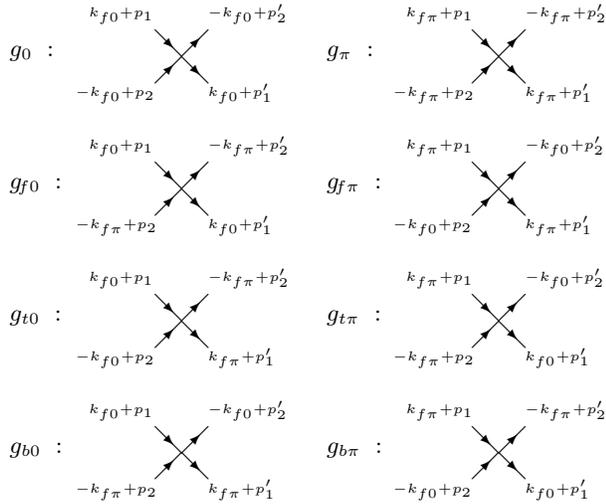
\begin{figure}[H]
\begin{center}
\newsavebox{\croix}
\begin{picture}(220,220)
\savebox{\croix}(20,20)[h]{%
\put(0,20){\line(1,-1){20}}
\put(0,0){\line(1,1){20}}
\put(7,13){\vector(1,-1){0}}
\put(17,3){\vector(1,-1){0}}
\put(7,7){\vector(1,1){0}}
\put(17,17){\vector(1,1){0}}}
\put(55,160){\usebox{\croix}}
\put(30,185){\makebox{$\scriptscriptstyle k_{f0}+p_1$}}
\put(25,155){\makebox{$\scriptscriptstyle -k_{f0}+p_2$}}
\put(75,185){\makebox{$\scriptscriptstyle -k_{f0}+p'_2$}}
\put(75,155){\makebox{$\scriptscriptstyle k_{f0}+p'_1$}}
\put(0,170){\makebox{$g_0\ :$}}

\put(55,110){\usebox{\croix}}
\put(30,135){\makebox{$\scriptscriptstyle k_{f0}+p_1$}}
\put(25,105){\makebox{$\scriptscriptstyle -k_{f\pi}+p_2$}}
\put(75,135){\makebox{$\scriptscriptstyle -k_{f\pi}+p'_2$}}
\put(75,105){\makebox{$\scriptscriptstyle k_{f0}+p'_1$}}
\put(0,120){\makebox{$g_{\!f0}\ :$}}

\put(55,60){\usebox{\croix}}
\put(30,85){\makebox{$\scriptscriptstyle k_{f0}+p_1$}}
\put(25,55){\makebox{$\scriptscriptstyle -k_{f0}+p_2$}}
\put(75,85){\makebox{$\scriptscriptstyle -k_{f\pi}+p'_2$}}
\put(75,55){\makebox{$\scriptscriptstyle k_{f\pi}+p'_1$}}
\put(0,70){\makebox{$g_{t0}\ :$}}

\put(55,10){\usebox{\croix}}
\put(30,35){\makebox{$\scriptscriptstyle k_{f0}+p_1$}}
\put(25,5){\makebox{$\scriptscriptstyle -k_{f\pi}+p_2$}}
\put(75,35){\makebox{$\scriptscriptstyle -k_{f0}+p'_2$}}
\put(75,5){\makebox{$\scriptscriptstyle k_{f\pi}+p'_1$}}
\put(0,20){\makebox{$g_{b0}\ :$}}

\put(175,160){\usebox{\croix}}
\put(150,185){\makebox{$\scriptscriptstyle k_{f\pi}+p_1$}}
\put(145,155){\makebox{$\scriptscriptstyle -k_{f\pi}+p_2$}}
\put(195,185){\makebox{$\scriptscriptstyle -k_{f\pi}+p'_2$}}
\put(195,155){\makebox{$\scriptscriptstyle k_{f\pi}+p'_1$}}
\put(120,170){\makebox{$g_\pi\ :$}}

\put(175,110){\usebox{\croix}}
\put(150,135){\makebox{$\scriptscriptstyle k_{f\pi}+p_1$}}
\put(145,105){\makebox{$\scriptscriptstyle -k_{f0}+p_2$}}
\put(195,135){\makebox{$\scriptscriptstyle -k_{f0}+p'_2$}}
\put(195,105){\makebox{$\scriptscriptstyle k_{f\pi}+p'_1$}}
\put(120,120){\makebox{$g_{\!f\pi}\ :$}}

\put(175,60){\usebox{\croix}}
\put(150,85){\makebox{$\scriptscriptstyle k_{f\pi}+p_1$}}
\put(145,55){\makebox{$\scriptscriptstyle -k_{f\pi}+p_2$}}
\put(195,85){\makebox{$\scriptscriptstyle -k_{f0}+p'_2$}}
\put(195,55){\makebox{$\scriptscriptstyle k_{f0}+p'_1$}}
\put(120,70){\makebox{$g_{t\pi}\ :$}}

\put(175,10){\usebox{\croix}}
\put(150,35){\makebox{$\scriptscriptstyle k_{f\pi}+p_1$}}
\put(145,5){\makebox{$\scriptscriptstyle -k_{f0}+p_2$}}
\put(195,35){\makebox{$\scriptscriptstyle -k_{f\pi}+p'_2$}}
\put(195,5){\makebox{$\scriptscriptstyle k_{f0}+p'_1$}}
\put(120,20){\makebox{$g_{b\pi}\ :$}}
\end{picture}
\end{center}
\caption{Schematic definitions of the couplings $g$}
\label{figcouplings}
\end{figure}

The relations between the different representations can be found in Refs.~%
\cite{Halboth} or \cite{Nickel}. Here, they reduce to~:
\renewcommand{\theequation}{\arabic{equation}}
\setcounter{equation}{3}
\begin{minipage}[h]{3.5cm}
\begin{equation}
\begin{array}{rcl}
g_s&=&-g_1-g_2\\g_t&=&g_1-g_2
\end{array}
\label{relgSC}
\end{equation}
\end{minipage}
\begin{minipage}[h]{3.5cm}
\begin{equation}
\begin{array}{rcl}
g_C&=&g_2-2g_1\\g_S&=&g_2
\end{array}
\label{relgSDW}
\end{equation}
\end{minipage}

\subsubsection{Other couplings}
\label{defcouplingsbis}

Here are the definitions of the different couplings $z$, from the couplings to
external fields $\cal Z$.

We omit the spin index $\alpha$ nor the symmetry index $\Gamma$, and
use the notation explained further in appendix~\ref{symbasic}. Mind
that $\Gamma=s,d,g\;$ for $\alpha=s$ (singlet) and $\Gamma=p,f$ for $\alpha=t$
(triplet). The symmetry ($s$, $d_{x^2-y^2}$, $g$, $p_x$, $f_x$ or
$f_y$) applying to each one is detailed in the main text.
$$
\begin{array}{c}
z^{\rm SC}_0(p_1,p_2,q)=\\
{\cal Z}^{\rm SC}(-k_{f0}+p_1,k_{f0}+p_2,(q,0),0)\\
z^{\rm SC}_\pi(p_1,p_2,q)=\\
{\cal Z}^{\rm SC}(-k_{f\pi}+p_1,k_{f\pi}+p_2,(q,0),\pi)\\
z^{\rm SC}_+(p_1,p_2,q)=\\
{\cal Z}^{\rm SC}(-k_{f\pi}+p_1,k_{f0}+p_2,(q,{\pi\over b}),0)\\
z^{\rm SC}_-(p_1,p_2,q)=\\
{\cal Z}^{\rm SC}(-k_{f0}+p_1,k_{f\pi}+p_2,(q,{\pi\over b}),\pi)\\
z^{\rm DW}_0(p_1,p_2,q)=\\
{\cal Z}^{\rm DW}(-k_{f0}+p_1,k_{f0}+p_2,(q-2k_{f0},0),0)\\
z^{\rm DW}_\pi(p_1,p_2,q)=\\
{\cal Z}^{\rm DW}(-k_{f\pi}+p_1,k_{f\pi}+p_2,(q-2k_{f\pi},0),\pi)\\
z^{\rm DW}_+(p_1,p_2,q)=\\
{\cal Z}^{\rm DW}(-k_{f\pi}+p_1,k_{f0}+p_2,(q-k_{f0}-k_{f\pi},{\pi\over b}),0)\\
z^{\rm DW}_-(p_1,p_2,q)=\\
{\cal Z}^{\rm DW}(-k_{f0}+p_1,k_{f\pi}+p_2,(q-k_{f0}-k_{f\pi},{\pi\over b}),\pi)
\end{array}
 $$

\vglue-1cm
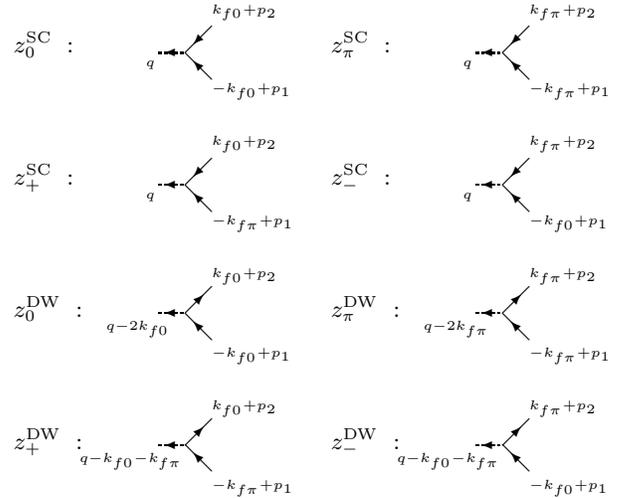
\begin{figure}[H]
\begin{center}
\newsavebox{\croisu}
\newsavebox{\croidw}
\begin{picture}(220,220)
\savebox{\croisu}(20,20)[h]{%
\put(20,20){\line(-1,-1){10}}
\put(20,0){\line(-1,1){10}}
\multiput(0,10)(2,0){5}{\line(1,0){1}}
\put(2,10){\vector(-1,0){0}}
\put(13,7){\vector(-1,1){0}}
\put(13,13){\vector(-1,-1){0}}}
\savebox{\croidw}(20,20)[h]{%
\put(20,20){\line(-1,-1){10}}
\put(20,0){\line(-1,1){10}}
\multiput(0,10)(2,0){5}{\line(1,0){1}}
\put(2,10){\vector(-1,0){0}}
\put(13,7){\vector(-1,1){0}}
\put(17,17){\vector(1,1){0}}}
\put(55,160){\usebox{\croisu}}
\put(50,165){\makebox{$\scriptscriptstyle q$}}
\put(75,185){\makebox{$\scriptscriptstyle k_{f0}+p_2$}}
\put(75,155){\makebox{$\scriptscriptstyle -k_{f0}+p_1$}}
\put(0,170){\makebox{$z^{\rm SC}_0~:$}}

\put(55,110){\usebox{\croisu}}
\put(50,115){\makebox{$\scriptscriptstyle q$}}
\put(75,135){\makebox{$\scriptscriptstyle k_{f0}+p_2$}}
\put(75,105){\makebox{$\scriptscriptstyle -k_{f\pi}+p_1$}}
\put(0,120){\makebox{$z^{\rm SC}_+~:$}}

\put(55,60){\usebox{\croidw}}
\put(35,65){\makebox{$\scriptscriptstyle q-2k_{f0}$}}
\put(75,85){\makebox{$\scriptscriptstyle k_{f0}+p_2$}}

\put(75,55){\makebox{$\scriptscriptstyle -k_{f0}+p_1$}}
\put(0,70){\makebox{$z^{\rm DW}_0\ :$}}

\put(55,10){\usebox{\croidw}}
\put(25,15){\makebox{$\scriptscriptstyle q-k_{f0}-k_{f\pi}$}}
\put(75,35){\makebox{$\scriptscriptstyle k_{f0}+p_2$}}
\put(75,5){\makebox{$\scriptscriptstyle -k_{f\pi}+p_1$}}
\put(0,20){\makebox{$z^{\rm DW}_+\ :$}}

\put(175,160){\usebox{\croisu}}
\put(170,165){\makebox{$\scriptscriptstyle q$}}
\put(195,185){\makebox{$\scriptscriptstyle k_{f\pi}+p_2$}}
\put(195,155){\makebox{$\scriptscriptstyle -k_{f\pi}+p_1$}}
\put(120,170){\makebox{$z^{\rm SC}_\pi~:$}}

\put(175,110){\usebox{\croisu}}
\put(170,115){\makebox{$\scriptscriptstyle q$}}
\put(195,135){\makebox{$\scriptscriptstyle k_{f\pi}+p_2$}}
\put(195,105){\makebox{$\scriptscriptstyle -k_{f0}+p_1$}}
\put(120,120){\makebox{$z^{\rm SC}_-~:$}}

\put(175,60){\usebox{\croidw}}
\put(155,65){\makebox{$\scriptscriptstyle q-2k_{f\pi}$}}
\put(195,85){\makebox{$\scriptscriptstyle k_{f\pi}+p_2$}}
\put(195,55){\makebox{$\scriptscriptstyle -k_{f\pi}+p_1$}}
\put(120,70){\makebox{$z^{\rm DW}_\pi\ :$}}

\put(175,10){\usebox{\croidw}}
\put(145,15){\makebox{$\scriptscriptstyle q-k_{f0}-k_{f\pi}$}}
\put(195,35){\makebox{$\scriptscriptstyle k_{f\pi}+p_2$}}
\put(195,5){\makebox{$\scriptscriptstyle -k_{f0}+p_1$}}
\put(120,20){\makebox{$z^{\rm DW}_-\ :$}}
\end{picture}
\end{center}
\caption{Schematic definitions of the couplings $z$}
\label{figcouplingsbis}
\end{figure}
\subsection{RG equations}

We give here the detailed RG equations.

\subsubsection{$g$ couplings}
\label{eqRGg}
\ \vglue0pt

Here are the RG equations for the couplings $g$, in $(c,l,p)$ representation.

The spin dependence is given, for all terms, by
$$
{dg_\alpha\over d\ell}=\sum_{\beta,\gamma}g_\beta\left({\cal C}^{\beta\gamma}_\alpha+
{\cal P}^{\beta\gamma}_\alpha\right)g_\gamma
 $$
where $\cal C$ and $\cal P$ correspond, respectively, to the Cooper and Peierls
channels, and are given, in the g-ology representation, by
$$
{\cal C}_1=-\pmatrix{0&1\cr1&0\cr}\quad
{\cal C}_2=-\pmatrix{1&0\cr0&1\cr}\quad
 $$
$$
{\cal P}_1=\pmatrix{2&-1\cr-1&0\cr}\quad
{\cal P}_2=\pmatrix{0&0\cr0&-1\cr}\ ,
 $$
see, for instance, Refs.~\cite{Zanchi,Halboth,Nickel}. In the following
equations, all two first terms are Cooper ones, whereas all two last
terms are Peierls ones; so, we omit the spin dependence, which is given
by the above equations, for each term. One gets
\begin{widetext}
$$
\begin{array}{l}
\qquad\qquad{\displaystyle{d\tilde g_0\over d\ell}}(c,l,p)=\\
{\Lambda\over8\Lambda+4|c|}\Big(\sum\limits_\pm
\tilde g_0({\scriptstyle c,\pm(\Lambda+{|c|\over2})+{l+p\over2},
\mp(\Lambda+{|c|\over2})+{l+p\over2}})
\tilde g_0({\scriptstyle c,\mp(\Lambda+{|c|\over2})+{l-p\over2},
\mp(\Lambda+{|c|\over2})-{l-p\over2}})\\
+\sum\limits_\pm \tilde g_{t0}({\scriptstyle c,\pm(\Lambda+{|c|\over2})+{l+p\over2},
\mp(\Lambda+{|c|\over2})+{l+p\over2}})
\tilde g_{t\pi}({\scriptstyle c,\mp(\Lambda+{|c|\over2})+{l-p\over2},
\mp(\Lambda+{|c|\over2})-{l-p\over2}})\Big)\\
+{\Lambda\over8\Lambda+4|p|}\Big(\sum\limits_\pm
\tilde g_0({\scriptstyle \mp(\Lambda+{|p|\over2})+{c+l\over2},
\pm(\Lambda+{|p|\over2})+{c+l\over2},p})
\tilde g_0({\scriptstyle \mp(\Lambda+{|p|\over2})+{c-l\over2},
\mp(\Lambda+{|p|\over2})-{c-l\over2},p})\Big)\\
+{\Lambda\over8\Lambda+4|p+2\Delta k_f|}\Big(\sum\limits_\pm
\tilde g_{b0}({\scriptstyle
\mp(\Lambda+{|p+2\Delta k_f|\over2})+{c+l\over2}-\Delta k_f,
\pm(\Lambda+{|p+2\Delta k_f|\over2})+{c+l\over2}-\Delta k_f,p})\\
\multicolumn{1}{r}{
\tilde g_{b\pi}({\scriptstyle
\mp(\Lambda+{|p+2\Delta k_f|\over2})+{c-l\over2}+\Delta k_f,
\mp(\Lambda+{|p+2\Delta k_f|\over2})-{c-l\over2}+\Delta k_f,p+2\Delta k_f})\Big)}
\end{array}
 $$

$$
\begin{array}{l}
\qquad\qquad{\displaystyle{d\tilde g_{\!f0}\over d\ell}}(c,l,p)=\\
{\Lambda\over8\Lambda+4|c|}\Big(\sum\limits_\pm
\tilde g_{\!f0}({\scriptstyle c,\pm(\Lambda+{|c|\over2})+{l+p\over2},
\mp(\Lambda+{|c|\over2})+{l+p\over2}})
\tilde g_{\!f0}({\scriptstyle c,\mp(\Lambda+{|c|\over2})+{l-p\over2},
\mp(\Lambda+{|c|\over2})-{l-p\over2}})\Big)\\
+{\Lambda\over8\Lambda+4|c+2\Delta k_f|}\Big(\sum\limits_\pm
\tilde g_{b0}({\scriptstyle c,
\pm(\Lambda+{|c+2\Delta k_f|\over2})+{l+p\over2}-\Delta k_f,
\mp(c\Lambda+{|c+2\Delta k_f|\over2})+{l+p\over2}-\Delta k_f})\\
\multicolumn{1}{r}{
\tilde g_{b\pi}({\scriptstyle c+2\Delta k_f,
\mp(\Lambda+{|c+2\Delta k_f|\over2})+{l-p\over2}+\Delta k_f,
\mp(\Lambda+{|c+2\Delta k_f|\over2})-{l-p\over2}}+\Delta k_f)\Big)}\\
+{\Lambda\over8\Lambda+4|p|}\Big(\sum\limits_\pm
\tilde g_{\!f0}({\scriptstyle \mp(\Lambda+{|p|\over2})+{c+l\over2},
\pm(\Lambda+{|p|\over2})+{c+l\over2},p})
\tilde g_{\!f0}({\scriptstyle \mp(\Lambda+{|p|\over2})+{c-l\over2},
\mp(\Lambda+{|p|\over2})-{c-l\over2},p})\\
+\sum\limits_\pm \tilde g_{t0}({\scriptstyle \mp(\Lambda+{|p|\over2})+{c+l\over2},
\pm(\Lambda+{|p|\over2})+{c+l\over2},p})
\tilde g_{t\pi}({\scriptstyle \mp(\Lambda+{|p|\over2})+{c-l\over2},
\mp(\Lambda+{|p|\over2})-{c-l\over2},p})\Big)
\end{array}
 $$

$$
\begin{array}{l}
\qquad\qquad{\displaystyle{d\tilde g_{t0}\over d\ell}}(c,l,p)=\\
{\Lambda\over8\Lambda+4|c|}\Big(\sum\limits_\pm
\tilde g_0({\scriptstyle c,\pm(\Lambda+{|c|\over2})+{l+p\over2},
\mp(\Lambda+{|c|\over2})+{l+p\over2}})
\tilde g_{t0}({\scriptstyle c,\mp(\Lambda+{|c|\over2})+{l-p\over2},
\mp(\Lambda+{|c|\over2})-{l-p\over2}})\\
+\sum\limits_\pm \tilde g_{t0}({\scriptstyle c,\pm(\Lambda+{|c|\over2})+{l+p\over2},
\mp(\Lambda+{|c|\over2})+{l+p\over2}})
\tilde g_\pi({\scriptstyle c,\mp(\Lambda+{|c|\over2})+{l-p\over2},
\mp(\Lambda+{|c|\over2})-{l-p\over2}})\Big)\\
+{\Lambda\over8\Lambda+4|p|}\Big(\sum\limits_\pm
\tilde g_{t0}({\scriptstyle \mp(\Lambda+{|p|\over2})+{c+l\over2},
\pm(\Lambda+{|p|\over2})+{c+l\over2},p})
\tilde g_{\!f\pi}({\scriptstyle \mp\Lambda+{|p|\over2})+{c-l\over2},
\mp(\Lambda+{|p|\over2})-{c-l\over2},p})\\
+\sum\limits_\pm \tilde g_{\!f0}({\scriptstyle \mp(\Lambda+{|p|\over2})+{c+l\over2},
\pm(\Lambda+{|p|\over2})+{c+l\over2},p})
\tilde g_{t0}({\scriptstyle \mp(\Lambda+{|p|\over2})+{c-l\over2},
\mp(\Lambda+{|p|\over2})-{c-l\over2},p})\Big)
\end{array}
 $$

$$
\begin{array}{l}
\qquad\qquad{\displaystyle{d\tilde g_{b0}\over d\ell}}(c,l,p)=\\
{\Lambda\over8\Lambda+4|c|}\Big(\sum\limits_\pm
\tilde g_{\!f\hspace{-1pt}0}({\scriptstyle c,\pm(\Lambda+{|c|\over2})+{l+p\over2}+\Delta k_f,
\mp(\Lambda+{|c|\over2})+{l+p\over2}+\Delta k_f})
\tilde g_{b0}({\scriptstyle c,\mp(\Lambda+{|c|\over2})+{l-p\over2}-\Delta k_f,
\mp(\Lambda+{|c|\over2})-{l-p\over2}-\Delta k_f})\Big)\\
+{\Lambda\over8\Lambda+4|c+2\Delta k_f|}\Big(\sum\limits_\pm
\tilde g_{b0}({\scriptstyle c,\pm(\Lambda+{|c+2\Delta k_f|\over2})+{l+p\over2},
\mp(\Lambda+{|c+2\Delta k_f|\over2})+{l+p\over2}})
\tilde g_{\!f\pi}({\scriptstyle c+2\Delta k_f,
\mp(\Lambda+{|c+2\Delta k_f|\over2}){l-p\over2},
\mp(\Lambda+{|c+2\Delta k_f|\over2})-{l-p\over2}})\Big)\\
+{\Lambda\over8\Lambda+4|p|}\Big(\sum\limits_\pm
\tilde g_0({\scriptstyle \mp(\Lambda+{|p|\over2})+{c+l\over2}+\Delta k_f,
\pm(\Lambda+{|p|\over2})+{c+l\over2}+\Delta k_f,p})
\tilde g_{b0}({\scriptstyle \mp(\Lambda+{|p|\over2})+{c-l\over2}-\Delta k_f,
\mp(\Lambda+{|p|\over2})-{c-l\over2}-\Delta k_f,p})\Big)\\
+{\Lambda\over8\Lambda+4|p+2\Delta k_f|}\Big(\sum\limits_\pm
\tilde g_{b0}({\scriptstyle \mp(\Lambda+{|p+2\Delta k_f|\over2})+{c+l\over2},
\pm(\Lambda+{|p+2\Delta k_f|\over2})+{c+l\over2},p})
\tilde g_\pi({\scriptstyle \mp(\Lambda+{|p+2\Delta k_f|\over2})+{c-l\over2},
\mp(\Lambda+{|p+2\Delta k_f|\over2})-{c-l\over2},p+2\Delta k_f})\Big)
\end{array}
 $$
\end{widetext}

\subsubsection{$z$ couplings}
\label{eqRGz}
\ \vglue0pt

Here are the RG equations for the couplings $z$, in $(k,c,p)$ representation.
The $z^{\rm SC}$ couplings should be written in the singlet/triplet
representation ($\alpha=s,t$), and the $z^{\rm DW}$ couplings should be written
in the Charge/Spin representation ($\alpha=C,S$). Then, the spin dependence
simply writes, for each one
$$
{dz_\alpha\over d\ell}=g_\alpha z_\alpha
 $$
and will therefore be again omitted. One gets
\begin{widetext}
$$
\begin{array}{l}
\qquad\qquad{\displaystyle{dz^{\rm SC}_0\over d\ell}}(c,k)=
{\Lambda\over4\Lambda+2|c|}\Big(\sum\limits_\pm
\tilde g_0({\scriptstyle c,\pm(\Lambda+{|c|\over2})+{c\over2}-k,
\mp(\Lambda+{|c|\over 2})+{c\over2}-k})
z^{\rm SC}_0({\scriptstyle c,\pm(\Lambda+{|c|\over2})+{c\over2}})\\
\multicolumn{1}{r}{
+\sum\limits_\pm\tilde g_{t0}({\scriptstyle c,
\pm(\Lambda+{|c|\over2})+{c\over2}-k,\mp(\Lambda+{|c|\over 2})+{c\over2}-k})
z^{\rm SC}_\pi({\scriptstyle c,\pm(\Lambda+{|c|\over2})+{c\over2}})\Big)}
\end{array}
 $$

$$
\begin{array}{l}
\qquad\qquad{\displaystyle{dz^{\rm SC}_+\over d\ell}}(c,k)=
{\Lambda\over4\Lambda+2|c|}\sum\limits_\pm
\tilde g_{f0}({\scriptstyle c,\pm(\Lambda+{|c|\over2})+{c\over2}-k,
\mp(\Lambda+{|c|\over 2})+{c\over2}-k})
z^{\rm SC}_+({\scriptstyle c,\pm(\Lambda+{|c|\over2})+{c\over2}})\\
+{\Lambda\over4\Lambda+2|c+2\Delta k_f|}\sum\limits_\pm
\tilde g_{b0}({\scriptstyle c,
\pm(\Lambda+{|c+2\Delta k_f|\over2})+{c\over2}-\Delta k_f-k,
\mp(\Lambda+{|c+2\Delta k_f|\over 2})+{c\over2}-\Delta k_f-k})
z^{\rm SC}_-({\scriptstyle c+2\Delta k_f,\pm(\Lambda+{|c+2\Delta k_f|\over2})
+{c\over2}+\Delta k_f})
\end{array}
 $$

$$
\begin{array}{l}
\qquad\qquad{\displaystyle{dz^{\rm DW}_0\over d\ell}}(p,k)=
{\Lambda\over4\Lambda+2|p|}\sum\limits_\pm
\tilde g_0({\scriptstyle \pm(\Lambda+{|p|\over2})+{p\over2}+k,
\pm(\Lambda+{|p|\over 2})-{p\over2}-k,p})
z^{\rm DW}_0({\scriptstyle p,\pm(\Lambda+{|p|\over2})-{p\over2}})\\
\!\!\!\!\!\!\!\!+{\Lambda\over4\Lambda+2|p+2\Delta k_f|}\sum\limits_\pm
\tilde g_{b\pi}({\scriptstyle
\pm(\Lambda+{|p+2\Delta k_f|\over2})+{p\over2}+\Delta k_f+k,
\pm(\Lambda+{|p+2\Delta k_f|\over 2})-{p\over2}+\Delta k_f-k,p+2\Delta k_f})
z^{\rm DW}_\pi({\scriptstyle p+2\Delta k_f,\pm(\Lambda+{|p+2\Delta k_f|\over2})
-{p\over2}-\Delta k_f})
\end{array}
 $$

$$
\begin{array}{l}
\qquad\qquad{\displaystyle{dz^{\rm DW}_+\over d\ell}}(p,k)=
{\Lambda\over4\Lambda+2|p|}\Big(\sum\limits_\pm
\tilde g_{f0}({\scriptstyle \pm(\Lambda+{|p|\over2})+{p\over2}+k,
\pm(\Lambda+{|p|\over 2})-{p\over2}-k,p})
z^{\rm DW}_+({\scriptstyle p,\pm(\Lambda+{|p|\over2})-{p\over2}})\\
\multicolumn{1}{r}{
+\sum\limits_\pm\tilde g_{t\pi}({\scriptstyle
\pm(\Lambda+{|p|\over2})+{p\over2}+k,\pm(\Lambda+{|p|\over 2})-{p\over2}-k,p})
z^{\rm DW}_-({\scriptstyle p,\pm(\Lambda+{|p|\over2})-{p\over2}})\Big)}
\end{array}
 $$

\subsubsection{$\chi$ couplings}
\label{eqRGc}
\ \vglue0pt

Here are the RG equations for the susceptibilities $\chi$, with the same
spin dependence as the corresponding $z$ couplings, which is again omitted,

$$
\begin{array}{l}
\qquad\qquad{\displaystyle{d\chi^{\rm SC}_0\over d\ell}}(q)=
-{\Lambda\over4\Lambda+2|q|}\Big(\sum\limits_\pm
z^{\rm SC}_0({\scriptstyle q,\pm(\Lambda+{|q|\over2})+{q\over2}})^2
+\sum\limits_\pm
z^{\rm SC}_\pi({\scriptstyle q,\pm(\Lambda+{|q|\over2})+{q\over2}})^2\Big)
\end{array}
 $$

$$
\begin{array}{l}
\qquad\qquad{\displaystyle{d\chi^{\rm SC}_+\over d\ell}}(q)=
-{\Lambda\over4\Lambda+2|q|}\sum\limits_\pm
z^{\rm SC}_+({\scriptstyle q,\pm(\Lambda+{|q|\over2})+{q\over2}})^2
-{\Lambda\over4\Lambda+2|q+2\Delta k_f|}\sum\limits_\pm
z^{\rm SC}_-({\scriptstyle q+2\Delta k_f,
\pm(\Lambda+{|q+2\Delta k_f|\over2})+{q\over2}+\Delta k_f})^2
\end{array}
 $$

$$
\begin{array}{l}
\!\!\!\!\!\!\qquad\qquad{\displaystyle{d\chi^{\rm DW}_0\over d\ell}}(q)=
-{\Lambda\over4\Lambda+2|q|}\sum\limits_\pm
z^{\rm DW}_0({\scriptstyle -q,\pm(\Lambda+{|q|\over2})+{q\over2}})^2
-{\Lambda\over4\Lambda+2|q-2\Delta k_f|}\sum\limits_\pm
z^{\rm DW}_\pi({\scriptstyle -q+2\Delta k_f,
\pm(\Lambda+{|q-2\Delta k_f|\over2})+{q\over2}-\Delta k_f})^2
\end{array}
 $$

$$
\begin{array}{l}
\qquad\qquad{\displaystyle{d\chi^{\rm DW}_+\over d\ell}}(q)=
-{\Lambda\over4\Lambda+2|q|}\Big(\sum\limits_\pm
z^{\rm DW}_+({\scriptstyle -q,\pm(\Lambda+{|q|\over2})+{q\over2}})^2
+\sum\limits_\pm
z^{\rm DW}_-({\scriptstyle -q,\pm(\Lambda+{|q|\over2})+{q\over2}})^2\Big)
\end{array}
 $$
\end{widetext}

\subsection{Symmetries}
\label{symmetries}
\subsubsection{Ordinary symmetries}
\label{symbasic}

If we apply the conjugation symmetry $C$ to the two-particle coupling $\cal G$,
we get~:
$$
{\cal G}(P'_1,P'_2,P_2,P_1)={\cal G}(-P_1,-P_2,-P'_2,-P'_1)\ ,
 $$
if we apply $A$, we get~:
$$
\begin{array}{c}
{\cal G}_C(P_2,P_1,P'_2,P'_1)=\\
   -2{\cal G}_C(P_1,P_2,P'_2,P'_1)-3{\cal G}_S(P_1,P_2,P'_2,P'_1)\\
{\cal G}_S(P_2,P_1,P'_2,P'_1)=\\
   {\cal G}_C(P_1,P_2,P'_2,P'_1)+2{\cal G}_S(P_1,P_2,P'_2,P'_1)\ ,
\end{array}
 $$
if we apply $A'$, we get~:
$$
\begin{array}{c}
{\cal G}_C(P_1,P_2,P'_1,P'_2)=\\
   -2{\cal G}_C(P_1,P_2,P'_2,P'_1)-3{\cal G}_S(P_1,P_2,P'_2,P'_1)\\
{\cal G}_S(P_1,P_2,P'_1,P'_2)=\\
   {\cal G}_C(P_1,P_2,P'_2,P'_1)+2{\cal G}_S(P_1,P_2,P'_2,P'_1)\ ,
\end{array}
 $$
and, finally, from parity $P$ conservation, we get
$$
{\cal G}(P_1,P_2,P'_2,P'_1)={\cal G}(-P_1,-P_2,-P'_2,-P'_1)\ .
 $$
Note that $AA'$ simply gives
${\cal G}(P_2,P_1,P'_1,P'_2)={\cal G}(P_1,P_2,P'_2,P'_1)$.

For the SC instability coupling, we will write the two-dimensional interaction
vector ${\bf Q}=(Q_\Vert,Q_\perp)$, and add a discrete variable
$\theta=0,\pi$, which indicates whether the $R$ particle is on the 0-band
($\theta=0$) or the $\pi$-band ($\theta=\pi$); this way, one can distinguish
0-0, 0-$\pi$, $\pi$-0 or $\pi$-$\pi$ processes (use Fig.~\ref{para2d} for help).

If we apply $P$, we get
$$
{\cal Z}^{\rm SC}(-P_1,-P_2,(-Q_\Vert,Q_\perp),\theta)
={\cal Z}^{\rm SC}(P_1,P_2,(Q_\Vert,Q_\perp),\theta)\ ,
 $$
if we apply $A$ or $A'$ (note that in $H_{\rm SC}$, the term with incoming
momenta $P_1$ and $P_2$ is conjugate to that with outgoing momenta $P_1$ and
$P_2$), we get
\begin{widetext}
$$
\begin{array}{rcl}
{\cal Z}^{\rm SC}_\alpha(P_2,P_1,(Q_\Vert,0),\theta)&=&
{\cal Z}^{\rm SC}_\alpha(P_1,P_2,(Q_\Vert,0),\theta) \quad \alpha=s,t\\
{\cal Z}^{{\rm SC}(s)}_s(P_2,P_1,(Q_\Vert,{\pi\over b}-Q_\perp),\theta)
&=&{\cal Z}^{{\rm SC}(s)}_s(P_1,P_2,(Q_\Vert,Q_\perp),\pi-\theta)\\
{\cal Z}^{{\rm SC}(g)}_s(P_2,P_1,(Q_\Vert,{\pi\over b}-Q_\perp),\theta)
&=&-{\cal Z}^{{\rm SC}(g)}_s(P_1,P_2,(Q_\Vert,Q_\perp),\pi-\theta)\\
{\cal Z}^{{\rm SC}(p_x)}_t(P_2,P_1,(Q_\Vert,{\pi\over b}-Q_\perp),\theta)&=&
{\cal Z}^{{\rm SC}(p_x)}_t(P_1,P_2,(Q_\Vert,Q_\perp),\pi-\theta)\\
{\cal Z}^{{\rm SC}(f_y)}_t(P_2,P_1,(Q_\Vert,{\pi\over b}-Q_\perp),\theta)&=&
-{\cal Z}^{{\rm SC}(f_y)}_t(P_1,P_2,(Q_\Vert,Q_\perp),\pi-\theta)\ .
\end{array}
 $$
\end{widetext}
Finally, it is interesting to note that ${\cal Z}^{\rm SC}_s$ (singlet) and
${\cal Z}^{\rm SC}_{t_x}$ (triplet) change sign under $S$ and are invariant
under $C$, while ${\cal Z}^{\rm SC}_{t_y}$ and ${\cal Z}^{\rm SC}_{t_z}$
(both triplet) do the opposite.

For the DW instability coupling, we use the same notation. Note that
$Q_\Vert$ writes \penalty-10000 $-(p_1-p_2+2k_{f\theta})$ for intraband
processes, and
$-(p_1-p_2+k_{f0}+k_{f\pi})$ for interband ones. If we apply $CS$, we get~:
$$
\begin{array}{c}
{\cal Z}^{\rm DW}(-P_1,-P_2,(2k_{f\theta}-Q_\Vert,Q_\perp),\theta)=\\
{\cal Z}^{\rm DW}(P_1,P_2,(-2k_{f\theta}+Q_\Vert,Q_\perp),\theta) \ ,
\end{array}
 $$
%\pagebreak[4]
and if we apply $AS$, we get
$$
\begin{array}{c}
{\cal Z}^{\rm DW}(P_2,P_1,(2k_{f\theta}-Q_\Vert,Q_\perp),\theta)=\\
\pm{\cal Z}^{\rm DW}(P_1,P_2,(-2k_{f\theta}+Q_\Vert,Q_\perp),\theta)
\end{array}
 $$
$$
\begin{array}{c}
{\cal Z}^{\rm DW}(P_2,P_1,
(k_{f0}+k_{f\pi}-Q_\Vert,{\pi\over b}-Q_\perp),\theta)=\\
\mp{\cal Z}^{\rm DW}(P_1,P_2,
(-k_{f0}-k_{f\pi}+Q_\Vert,Q_\perp),\pi-\theta) \ ,
\end{array}
 $$
where $\pm$ reads $+$ for site ordering, and $-$ for bond ordering.

\subsubsection{Supplementary symmetry}
\label{symsupp}

When we apply the special symmetry $\tilde C$ to two-particle couplings
$\cal G$, we get~:
\begin{widetext}
$$
{\cal G}(k_{f0}+k_{f\pi}-P_1,k_{f0}+k_{f\pi}-P_2,
k_{f0}+k_{f\pi}-P'_2,k_{f0}+k_{f\pi}-P'_1)=
{\cal G}(P_1,P_2,P'_2,P'_1)
 $$

When we apply the special symmetry $\tilde C$ to SC instabilities
${\cal Z}_{\rm SC}$, we get~:
\begin{eqnarray*}
{\cal Z}^{\rm SC(\Gamma)}_\alpha(-k_{f0}-k_{f\pi}-P_1,k_{f0}+k_{f\pi}-P_2,
(-Q_\Vert,0),\theta)
=\pm{\cal Z}^{\rm SC(\Gamma)}_\alpha
(P_1,P_2,(Q_\Vert,0),\pi-\theta)\\
{\cal Z}^{\rm SC(\Gamma)}_\alpha(-k_{f0}-k_{f\pi}-P_1,k_{f0}+k_{f\pi}-P_2,
(-Q_\Vert,{\pi\over b}-Q_\perp),\theta)=
\pm{\cal Z}^{\rm SC(\Gamma)}_\alpha
(P_1,P_2,(Q_\Vert,Q_\perp),\pi-\theta)
\end{eqnarray*}
where $\pm$ reads + for $\alpha=s$, $\Gamma=s$ or for $\alpha=t$, $\Gamma=p_x$,
and $-$ for $\alpha=s$,  $\Gamma=d,g$ or for $\alpha=t$, $\Gamma=f_x,f_y$.

When we apply the special symmetry $\tilde C$ to DW instabilities
${\cal Z}_{\rm DW}$, we get~:
$$
{\cal Z}^{\rm DW}(-k_{f0}-k_{f\pi}-P_1,k_{f0}+k_{f\pi}-P_2,
(2k_{f\theta}-Q_\Vert,0),\theta)=
\pm{\cal Z}^{\rm DW}(P_1,P_2,
(-2k_{f[\pi-\theta]}+Q_\Vert,0),\pi-\theta)
 $$%\small
$$
{\cal Z}^{\rm DW}(-k_{f0}-k_{f\pi}-P_1,k_{f0}+k_{f\pi}-P_2,
(-k_{f0}-k_{f\pi}-Q_\Vert,{\pi\over b}-Q_\perp),\theta)=
\pm{\cal Z}^{\rm DW}(P_1,P_2,
(-k_{f0}-k_{f\pi}+Q_\Vert,Q_\perp),\pi-\theta)
 $$
where $\pm$ reads $+$ for site ordering, and $-$ for bond ordering.

\subsubsection{$g_b$ orbits}
\label{orbites}

Here are the 8 first orbits of the $g_{b0}$ coefficient, in $(c,l,p)$
representation~:

{
\begin{eqnarray*}
&&\{(0,0,-2\Delta k_f,0),(0,-2\Delta k_f,0),
(-2\Delta k_f,0,0),(-2\Delta k_f,-2\Delta k_f,-2\Delta k_f)\}
\mbox{ are sym. equiv.}\\
&&\{(-4\Delta k_f,-4\Delta k_f,-2\Delta k_f),(-4\Delta k_f,2\Delta k_f,0),
(2\Delta k_f,-4\Delta k_f,0),(2\Delta k_f,2\Delta k_f,-2\Delta k_f)\}
\mbox{ id.}\\
&&\{(0,-4\Delta k_f,2\Delta k_f),(0,2\Delta k_f,-4\Delta k_f),
(-2\Delta k_f,-4\Delta k_f,-4\Delta k_f),(-2\Delta k_f,2\Delta k_f,2\Delta k_f)\}
\mbox{ id.}\\
&&\{(-4\Delta k_f,0,2\Delta k_f),(-4\Delta k_f,-2\Delta k_f,-4\Delta k_f),
(2\Delta k_f,0,-4\Delta k_f),(2\Delta k_f,-2\Delta k_f,2\Delta k_f)\}
\mbox{ id.}\\
&&\{(0,-4\Delta k_f,-2\Delta k_f),(0,2\Delta k_f,0),
(-2\Delta k_f,-4\Delta k_f,0),(-2\Delta k_f,2\Delta k_f,-2\Delta k_f)\}
\mbox{ id.}\\
&&\{(-4\Delta k_f,0,-2\Delta k_f),(-4\Delta k_f,-2\Delta k_f,0),
(2\Delta k_f,0,0),(2\Delta k_f,-2\Delta k_f,-2\Delta k_f)\}
\mbox{ id.}\\
&&\{(0,0,2\Delta k_f),(0,-2\Delta k_f,-4\Delta k_f),
(-2\Delta k_f,0,-4\Delta k_f),(-2\Delta k_f,-2\Delta k_f,2\Delta k_f)\}
\mbox{ id.}\\
&&\{(-4\Delta k_f,-4\Delta k_f,2\Delta k_f),(-4\Delta k_f,2\Delta k_f,-4\Delta k_f),
(2\Delta k_f,-4\Delta k_f,-4\Delta k_f),(2\Delta k_f,2\Delta k_f,2\Delta k_f)\}
\mbox{ id.}
\end{eqnarray*}}
\end{widetext}

\subsection{Fourier Transform}
\label{Fourier}

\paragraph*{Creation and annihilation operators}

If one writes $\psi_{ij\sigma}^\dag$ the creator of a particle of spin
$\sigma$, located in real space at position $i$ ($i\in\{1,\cdots,N\}$), on
chain $j$ ($j=\pm1$), the representation in the momentum space writes
\begin{eqnarray*}
L_{p,0,\sigma}=\Psi_{(-k_{f0}+p,0)\sigma}&=&
\sum_{ij}\psi_{ij\sigma}\e^{-\ii(p-k_{f0})ia}\\
L_{p,\pi,\sigma}=\Psi_{(-k_{f\pi}+p,{\pi\over b})\sigma}&=&
\sum_{ij}j\psi_{ij\sigma}\e^{-\ii(p-k_{f\pi})ia}\\
R_{p,0,\sigma}=\Psi_{(k_{f0}+p,0)\sigma}&=&
\sum_{ij}\psi_{ij\sigma}\e^{-\ii(p+k_{f0})ia}\\
R_{p,\pi,\sigma}=\Psi_{(k_{f\pi}+p,{\pi\over b})\sigma}&=&
\sum_{ij}j\psi_{ij\sigma}\e^{-\ii(p+k_{f\pi})ia}
\end{eqnarray*}
with the notations of the text. $\Psi_{\bf k\sigma}$ stands for the absolute
momentum representation, while $L$ and $R$ stand for the relative momentum
representation. These relations are given for annihilation operators, one must
take the complex conjugation to obtain those for the creation operators.

The reverse relations simply write, in terms of the $\Psi$ operators,
$$
\psi_{ij\sigma}=\int_{-{\pi\over a}}^{\pi\over a}{adP\over4\pi}
\e^{\ii aip}(\Psi_{(P,0),\sigma}+j\Psi_{(P,{\pi\over b}),\sigma})\ ;
 $$
but one can also express them in terms of the $L$ and $R$ operators. Then,
one can check that this transformation is the inverse of the first one.

\paragraph*{Electron-electron pair operator}

In real space, the SC order parameters are the mean value of the
electron-electron pair operator, which writes
$$
O^\alpha({\bf X})=\sum_{\bf X'\atop\sigma\sigma'}
\psi^{\phantom{\dag}}_{\bf X\sigma}
\psi^{\phantom{\dag}}_{\bf X'\sigma'}\Gamma({\bf X,X'})
\tau^\alpha_{\sigma\sigma'}\ .
 $$

To each ${\bf Q}=(Q_\Vert,Q_\perp)$ corresponds a Fourier component
$$
O^\alpha({\bf Q})=\sum_{\bf X}\e^{-\bf Q.X}O^\alpha({\bf X})=
\sum_{ij}\e^{-\ii(aQ_\Vert i+bQ_\perp j/2)}O^\alpha(\bf X)\ .
 $$

We only keep the componants $\bf Q$ which lead to singularities; as explained in
the main text, they are $(0,0)$ and $(\pm\Delta k_f,{\pi\over b})$. So, using
the short notation $0$ for the first and $\pi_\pm$ for the second, one gets
$$
O^\alpha(0)=\sum_iO^\alpha(ai,1)+O^\alpha(ai,-1)
 $$
and
$$
O^\alpha(\pi_\pm)=\sum_i-\ii\e^{\mp\ii\Delta k_f a}
\big(O^\alpha(ai,1)-O^\alpha(ai,-1)\big)
 $$
where the main factor
$
O^\alpha(ai,1)\pm O^\alpha(ai,-1)=
\sum_{jj'\atop\sigma\sigma'}(\pm1)^j\psi_{ij\sigma}\phi_{i'j'\sigma'}
\Gamma(a(i-i'),b(j-j')/2)\tau^\alpha_{\sigma\sigma'}
$
is the mixed representation of the pair operator\cite{Poilblanc}.

Eventually, the $\psi_{ij\sigma}$ can be expressed in terms of the
$\Psi_{\bf P\sigma}$, so that the componants write
$$
O^\alpha({\bf Q})=\int_{-\pi\over a}^{\pi\over a}{adP\over4\pi}
\sum_{\theta=0,\pi}z_{\bf Q}({\bf p})
\sum_{\sigma\sigma'}
\Psi_{\bf p,\sigma}\Psi_{\bf Q-p,\sigma'}\tau^\alpha_{\sigma\sigma'}
 $$
where ${\bf p}=(P,{\theta\over b})$ and we will also use
${\bf Q}=(Q_\Vert,Q_\perp)$. Be careful that, for instance, with
${\bf Q}=(\Delta k_f,{\pi\over b})$ and ${\bf p}=(p-k_{f0},0)$, and thus
$\Psi_{\bf p,\sigma}=L_{p,0\sigma}$, the calculation of $\Psi_{\bf Q-p,\sigma'}$
is not immediate; one gets
$\Psi_{\bf Q-p,\sigma'}=\Psi_{(\Delta k_f-p+k_{f0},{\pi\over b}),\sigma'}
=R_{2\Delta k_f-p,\pi,\sigma'}$.

With $\Gamma({\bf X,X'})=\delta_{ii'}\delta_{jj'}$, one finds $z_0({\bf p})=1$
(singlet 0-condensate of $s$ symmetry), and $z_{\pi_\pm}({\bf p})=-\ii$ (singlet
$\pi$-condensate of $s$ symmetry). With $\Gamma({\bf X,X'})=
\delta_{ii'}\delta_{j,-j'}$, one finds $z_0({\bf p})=\cos(\theta)$ (singlet
0-condensate of $d$ symmetry) and $z_{\pi_\pm}({\bf p})=\ii\cos(\theta)$
(triplet $\pi$-condensate of $f_y$ symmetry). With $\Gamma({\bf X,X'})=
\delta_{i,i'\mp1}\delta_{j,-j'}$, one finds $z_0({\bf p})=\cos(aP)\cos(\theta)$
(singlet 0-condensate of extended $d$ symmetry), as well as
$z_0({\bf p})=-\ii\sin(aP)\cos(\theta)$ (triplet 0-condensate of $f_x$ symmetry),
and $z_{\pi_\pm}({\bf p})=\sin(a(P\mp{\Delta k_f\over2}))\cos(\theta)$
(singlet $\pi$-condensate of $g$ symmetry) or $z_{\pi_\pm}({\bf p})=
\ii\e^{\pm\ii{\Delta k_fa\over2}}\cos(a(P\mp{\Delta k_f\over2}))\cos(\theta)$
(triplet $\pi$-condensate of extended $f_y$ symmetry). With

$\Gamma({\bf X,X'})=\delta_{i,i'\mp1}\delta_{jj'}$, one finds
$z_0({\bf p})=\cos(aP)$ (singlet 0-condensate of extended $s$ symmetry) or
$z_0({\bf p})=-\ii\sin(aP)$ (triplet 0-condensate of $p_x$ symmetry), and
$z_{\pi_\pm}({\bf p})=-\ii\e^{\pm\ii{\Delta k_{\!f}a\over2}}
\cos(a(P\mp{\Delta k_f\over2}))$
(singlet $\pi$-condensate of extended $s$ symmetry) or $z_{\pi_\pm}({\bf p})=
-\e^{\pm\ii{\Delta k_fa\over2}}\sin(a(P\mp{\Delta k_f\over2}))$
(triplet $\pi$-condensate of $p_x$ symmetry).

\paragraph*{Electron-hole pair operator}

It is almost the same, with the product of a creation and an annihilation
operators, be careful, however, that, in reciprocal space, one gets~:
$$
\sum_{\sigma\sigma'}\int_{-\pi\over a}^{\pi\over a}{adP\over4\pi}
\Psi^\dag_{\bf p \sigma}\Psi_{\bf Q+p\sigma'}^{\phantom{\dag}}z({\bf p})
\tau^\alpha_{\sigma\sigma'} \ .
 $$

\end{document}